\documentclass[aps,prx,twocolumn,reprint,longbibliography,superscriptaddress]{revtex4-1}

\usepackage{graphicx}
\usepackage{bm}
\usepackage{amssymb}
\usepackage{amsmath}

\begin{document}

\title{Switching of Charge-Current-Induced Spin Polarization \\
in the Topological Insulator BiSbTeSe$_2$}

\author{Fan Yang}
\email[Electronic mail: ]{yang@ph2.uni-koeln.de}
\affiliation{II. Physikalisches Institut, Universit\"{a}t zu K\"{o}ln, Z\"{u}lpicher Stra\ss{}e 77, 50937 K\"{o}ln, Germany}
\affiliation{Institute of Scientific and Industrial Research, Osaka University, Ibaraki, Osaka 567-0047, Japan}

\author{Subhamoy Ghatak}
\affiliation{II. Physikalisches Institut, Universit\"{a}t zu K\"{o}ln, Z\"{u}lpicher Stra\ss{}e 77, 50937 K\"{o}ln, Germany}
\affiliation{Institute of Scientific and Industrial Research, Osaka University, Ibaraki, Osaka 567-0047, Japan}

\author{A. A. Taskin}
\affiliation{II. Physikalisches Institut, Universit\"{a}t zu K\"{o} ln, Z\"{u}lpicher Stra\ss{}e 77, 50937 K\"{o}ln, Germany}
\affiliation{Institute of Scientific and Industrial Research, Osaka University, Ibaraki, Osaka 567-0047, Japan}

\author{Kouji Segawa}
\affiliation{Institute of Scientific and Industrial Research, Osaka University, Ibaraki, Osaka 567-0047, Japan}
\affiliation{Department of Physics, Kyoto Sangyo University, Kyoto 603-8555 Japan}

\author{Yuichiro~Ando}
\affiliation{Department of Electronic Science and Engineering, Kyoto University, Kyoto 615-8510, Japan}

\author{Masashi~Shiraishi}
\affiliation{Department of Electronic Science and Engineering, Kyoto University, Kyoto 615-8510, Japan}

\author{Yasushi Kanai}
\affiliation{Institute of Scientific and Industrial Research, Osaka University, Ibaraki, Osaka 567-0047, Japan}

\author{Kazuhiko Matsumoto}
\affiliation{Institute of Scientific and Industrial Research, Osaka University, Ibaraki, Osaka 567-0047, Japan}

\author{Achim Rosch}
\affiliation{Institut f\"{u}r Theoretische Physik, Universit\"{a}t zu K\"{o}ln, Z\"{u}lpicher Stra\ss{}e 77, 50937 K\"{o}ln, Germany}

\author{Yoichi Ando}
\email[Electronic mail: ]{ando@ph2.uni-koeln.de}
\affiliation{II. Physikalisches Institut, Universit\"{a}t zu K\"{o}ln, Z\"{u}lpicher Stra\ss{}e 77, 50937 K\"{o}ln, Germany}
\affiliation{Institute of Scientific and Industrial Research, Osaka University, Ibaraki, Osaka 567-0047, Japan}

\date{\today}

\begin{abstract}

The charge-current-induced spin polarization is a key property of topological insulators for their applications in spintronics. However, topological surface states are expected to give rise to only one type of spin polarization for a given current direction, which has been a limiting factor for spin manipulations. Here we report that in devices based on the bulk-insulating topological insulator BiSbTeSe$_2$, an unexpected switching of spin polarization was observed upon changing the chemical potential. The spin polarization expected from the topological surface states was detected in a heavily electron-doped device, whereas the opposite polarization was reproducibly observed in devices with low carrier densities. We propose that the latter type of spin polarization stems from topologically-trivial two-dimensional states with a large Rashba spin splitting, which are caused by a strong band bending at the surface of BiSbTeSe$_2$ beneath the ferromagnetic electrode used as a spin detector. This finding paves the way for realizing the ``spin transistor" operation in future topological spintronic devices.

\end{abstract}

\pacs{72.25.-b, 73.20.At, 85.75.-d, 75.47.-m}

\maketitle

\section{Introduction}

The surface states of three-dimensional (3D) topological insulators (TIs) \cite{TI-review-Hasan, TI-review-Qi, TI-review-Ando} possess a helical spin texture in which the spin and momentum are perpendicularly locked to each other. Due to this spin-momentum locking, a net spin polarization can be induced by a charge current and \textit{vice versa}. This peculiar property makes 3D TIs a promising platform for spintronic applications \cite{TS-theory-Burkov, TS-theory-Pesin, TS-theory-Mahfouzi}, and as a result, there is growing interest in topological spintronics. Already a number of experiments with various approaches have been carried out on 3D TIs
\cite{spin-torque-Mellnik, spin-torque-Fan, spin-torque-Wang, spin-pumping-Shiomi, spin-pumping-Jamali, spin-pumping-Deorani, spin-pumping-Shiomi, ele-dection-Li, ele-dection-Dankert, ele-dection-Tang, ele-dection-Liu, ele-dection-Lee, ele-dection-Tian, ele-dection-Ando,ele-detection-Vaklinova}, including spin-transfer torque \cite{spin-torque-Mellnik, spin-torque-Fan, spin-torque-Wang}, spin pumping \cite{spin-pumping-Shiomi, spin-pumping-Jamali, spin-pumping-Deorani}, and all-electrical measurement \cite{ele-dection-Li, ele-dection-Ando, ele-dection-Dankert, ele-dection-Tang, ele-dection-Liu, ele-dection-Lee, ele-dection-Tian,ele-detection-Vaklinova}. Among these experiments, the all-electrical measurement of the charge-current-induced spin polarization is of particular importance due to its direct applicability to spintronics.

For spintronic applications, it is important to be able to perform ``spin-transistor" operation, that is, to switch the orientation of the charge-current-induced spin polarization via electrostatic gating. In this regard, the switching of the spin helicity across the Dirac point in the topological surface state (TSS) of a 3D TI may seem promising at first sight, but this expectation is in fact ungrounded, because the Fermi velocity of electrons with a given spin orientation does not change across the Dirac point (detailed explanations of this point is given later). Nevertheless, in the literature two different types of current-induced spin polarizations have been observed. So far both types of results have been claimed to be due to the TSS, leaving the situation controversial. If this controversy is sorted out and the origin of the two types of current-induced spin polarizations is understood, such an understanding may make it possible to conceive a spin transistor based on 3D TIs. Here, we directly address this problem and offer a possible solution.

The electrical detection of the current-induced spin polarization has been reported in various 3D TI materials, including Bi$_2$Se$_3$ \cite{ele-dection-Li, ele-dection-Dankert}, Bi$_{1.5}$Sb$_{0.5}$Te$_{1.7}$Se$_{1.3}$ \cite{ele-dection-Ando}, (Bi$_{1-x}$Sb$_x$)$_2$Te$_3$ \cite{ele-dection-Tang, ele-dection-Liu, ele-dection-Lee}, and Bi$_2$Te$_2$Se \cite{ele-dection-Tian}. Most experiments use a ferromagnetic contact as a spin detector and measure spin-dependent voltage between the ferromagnet (FM) and TI. However, conflicting interpretations of the spin voltage $V_{\bm{S}}$ have been claimed in the literature. Specifically, Li {\it et al.} \cite{ele-dection-Li} attributed positive $V_{\bm{S}}$ to the situation when the magnetization of the FM detector $\bm{M}_{\rm{FM}}$ is antiparallel to the induced spin polarization $\bm{S}$ (i.e. $\bm{M}_{\rm{FM}} \parallel -\bm{S}$), whereas most other papers \cite{ele-dection-Dankert, ele-dection-Tang, ele-dection-Lee, ele-dection-Liu, ele-dection-Tian} attribute positive $V_{\bm{S}}$ to the situation $\bm{M}_{\rm{FM}} \parallel \bm{S}$.
Theoretically \cite{CIS-model-Hong}, $V_{\bm{S}}$ should be positive for $M_{\rm{FM}}\parallel \bm{S}$. Furthermore, as long as the TSS is responsible for the induced $\bm{S}$, the orientation of $\bm{S}$ is expected to be the same for both $n$-type and $p$-type carriers \cite{CIS-model-Hong}.
Indeed, the expected polarity was observed in a recent experiment in which the carrier type of a TI device was continuously changed from $n$-type to $p$-type by backgating \cite{ele-dection-Lee}.

Hence, the contribution of the TSS to the spin voltage has become largely understood, but the reason of the controversial observation by Li {\it et al.} \cite{ele-dection-Li} remains a puzzle. In this respect, it is useful to notice that the possible contribution of the topologically-trivial two-dimensional (2D) states with a large Rashba spin splitting, which is caused by surface band bending and often coexist with the TSS \cite{Rashba-Bianchi, Rashba-King, Rashba-Zhu, Rashba-Bahramy}, has not been well understood. Such Rashba states could lead to an opposite sign of the current-induced spin polarization \cite{CIS-model-Hong}. One should note that any difference in the work functions between FM and TI would lead to a band bending at the interface and could cause a pronounced effect in spin detection, but this possibility has been neglected in previous works \cite{ele-dection-Li, ele-dection-Dankert, ele-dection-Tang, ele-dection-Lee, ele-dection-Liu, ele-dection-Tian, ele-dection-Ando}.

In this paper, we report the current-induced spin polarization in devices made from the topological insulator BiSbTeSe$_2$. In the as-grown crystals of this material, the chemical potential lies slightly below the Dirac point of the surface state, and its position changes easily when the sample is annealed; this makes it possible for us to investigate spin-voltage devices with different carrier types and densities. In Sec. II, we will present a transparent review of the physics behind the spin-dependent voltage in the TI-based devices and discuss the correct sign of the current-induced spin polarization generated in the TSS.
We will then present our experiment, in which two different types of spin polarization were detected. The first type was observed in a strongly $n$-type doped BiSbTeSe$_2$ device and its polarity meets the expectation for the TSS. The second type was observed in two low-carrier-density devices and their spin polarization cannot be ascribed to the TSS. We propose that the latter originates from the topologically-trivial 2D states with Rashba spin splitting, and discuss why they are expected in low-carrier-density samples. The possibility to utilize the Rashba-split 2D states alongside the TSS opens new pathways in future topological spintronic devices, in particular the prospect for spin-transistor operations.

\section{Principle of spin detection}

\subsection{Sign of the spin-dependent voltage}

\begin{figure*}
\includegraphics[width=0.9 \linewidth]{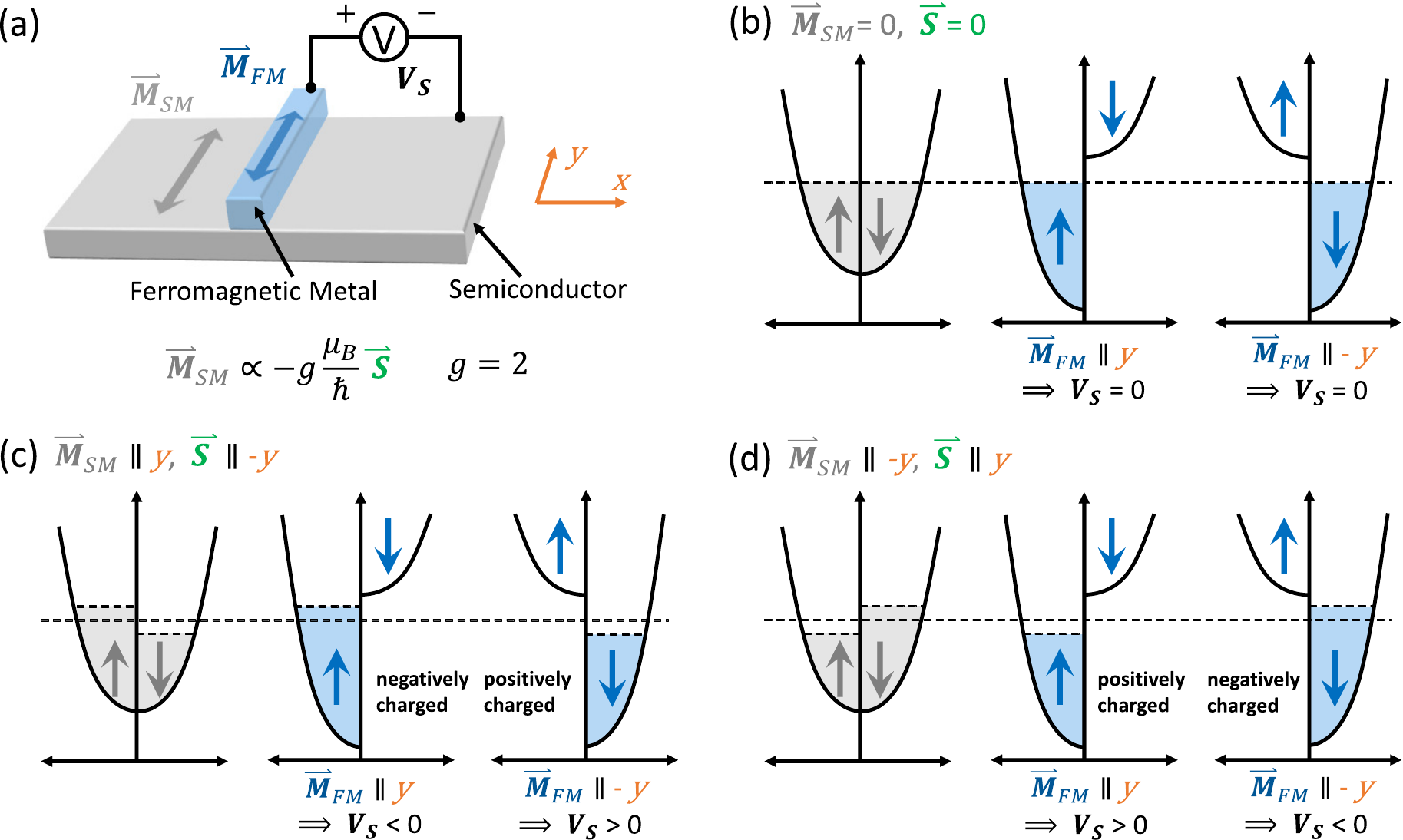}
\caption{\label{fig:Fig1} {
Principle of spin detection. The arrows in this figure denote the direction of {\it magnetization}, which is opposite to the direction of spin polarization, as explained in the text. (a) Measurement configuration for detecting the nonequilibrium spin polarization. A ferromagnetic tunneling contact fabricated on top of a semiconductor is used as a spin detector. (b)-(d) Illustrations to demonstrate the sign of the spin-dependent voltage. In the presence of a nonequilibrium spin polarization, the electrochemical potential of the semiconductor becomes magnetization-dependent. Due to the electron exchange, the chemical potential of the ferromagnet always aligns with that of the sub-band with the same magnetization direction, which makes the ferromagnet positively or negatively charged and gives rise to a positive or negative spin-dependent voltage on the ferromagnet.
}}
\end{figure*}

It has been established that the non-equilibrium spin polarization can be detected by using a FM contact as a spin detector. Early theories and experiments of spin detection started from the 1980s \cite{SpinVoltage-Johnson-01, SpinVoltage-Johnson-02}. At that time, ohmic contacts between FM and a normal metal were employed for spin detection. However, it eventually turned out that ohmic contacts are not efficient for detecting the spin polarization occurring in semiconductors due to the so-called conductance mismatch \cite{SpinDetectTheo-Schmidt}: When the impedances of the FM and the material beneath it are significantly different, the spin-dependent voltage becomes too small to be measured. In 2000, it was theoretically proposed that the problem of conductance mismatch can be solved by inserting a tunnel barrier between the FM detector and the semiconductor \cite{SpinDetectTheo-Rashba}. Corresponding experimental techniques were developed several years later \cite{SpinDetectExp-Lou} and they are recently applied to detect the current-induced spin polarization in TIs \cite{ele-dection-Li, ele-dection-Dankert, ele-dection-Tang, ele-dection-Liu, ele-dection-Lee, ele-dection-Tian}.

As already mentioned, there is a controversy over the sign of the spin-dependent voltage in the previous studies on TIs \cite{ele-dection-Li, ele-dection-Dankert, ele-dection-Tang, ele-dection-Lee, ele-dection-Liu, ele-dection-Tian, ele-dection-Ando}. This controversy comes down to the question, ``What is the proper sign of $V_{\bm{S}}$ for $\bm{M}_{\rm{FM}} \parallel \bm{S}$?" Here $V_{\bm{S}}$ and $\bm{M}_{\rm{FM}}$ denote the spin-dependent voltage and the magnetization in the FM spin detector, and $\bm{S}$ is the non-equilibrium spin polarization to be detected.
In the first paper by Li {\it et al.} \cite{ele-dection-Li}, it was argued that \textit{negative} $V_{\rm{S}}$ corresponds to $\bm{M}_{\rm{FM}} \parallel \bm{S}$, whereas in the other papers \cite{ele-dection-Dankert, ele-dection-Tang, ele-dection-Liu, ele-dection-Tian}, the opposite was claimed to be true, that is, \textit{positive} $V_{\rm{S}}$ corresponds to $\bm{M}_{\rm{FM}}\parallel\bm{S}$. In the following, we review the basic principles of spin detection in the case of TIs and explain from the ground up why the latter is correct. This clarification forms the basis of our interpretation of the two different types of $V_{\bm{S}}$ signals observed in BiSbTeSe$_2$

The principle of spin detection is illustrated in Fig. 1. The key idea is that, due to the electron exchange between the FM and the semiconductor, the FM always equilibrate with the electrochemical potential of the subband which has the same magnetization direction. To be concrete, we consider the basic configuration shown in Fig. 1(a). A FM contact with a tunnel barrier, which works as a spin detector, is placed on top of a semiconductor (SM). For simplicity, we assume that the FM is a half-metal, namely, only one subband is at the Fermi level. The voltage $V_{\bm{S}}$ is measured on the FM with respect to the SM. Hereafter, we sometimes use $\uparrow$ and $\downarrow$ to denote the $y$ and $-y$ directions, respectively.

It is important to notice that, due to the negative charge, the electron magnetic moment $\bm{\mu}_e$ is {\it antiparallel} to the spin vector $\bm{s}$, i.e., $\bm{\mu}_e = -(g\mu _{B}/\hbar)\bm{s}$, where $g$ is the $g$-factor and $\mu_B$ is the Bohr magneton. This means that the $\bm{M}_{\rm{SM}\uparrow}$ subband corresponds to the $\bm{s}_{\downarrow}$ subband, and the $\bm{M}_{\rm{SM}\downarrow}$ subband corresponds to the $\bm{s}_{\uparrow}$ subband. Note that all the arrows in the band diagrams in Fig. 1 denote the direction of the \textit{magnetization}, which is opposite to the direction of spin polarization.

In the absence of spin polarization in the SM, the electrochemical potentials of $\bm{M}_{\rm{SM}\uparrow}$ and $\bm{M}_{\rm{SM}\downarrow}$ subbands are equal. The chemical potential of the ferromagnet is also at the same level, irrespective of its magnetization direction. In this case, the FM is electrically neutral and hence $V_{\bm{S}}=0$, as shown in Fig. 2(b).

When a non-equilibrium spin polarization $\bm{S}$ is induced in the SM via methods like spin injection \cite{SpinVoltage-Johnson-01,SpinVoltage-Johnson-02,SpinDetectExp-Lou}, a corresponding magnetization $\bm{M}_{\rm{SM}}$ is generated in the $\bm{-S}$ direction. The non-zero $\bm{M}_{\rm{SM}}$ leads to an increase (decrease) in the electrochemical potential of the majority (minority) subband, with the potential difference $\Delta \mu = \mu_{\rm maj}-\mu_{\rm min} > 0$ (``maj'' and ``min'' stand for majority and minority, respectively). When $\bm{M}_{\rm{FM}} \parallel \bm{M}_{\rm{SM}}$, electrons in the majority subband of the SM will move into the FM and raise its potential $\mu_{\rm{FM}}$, until $\mu_{\rm{FM}} = \mu_{\rm maj}$ is reached. In this process, the FM become negatively charged, giving rise to a negative voltage $V_{\bm{S}}=(\Delta \mu/2)/(-e)=-\Delta \mu/(2e)< 0$. On the other hand, when $\bm{M}_{\rm{FM}} \parallel \bm{-M}_{\rm{SM}}$, electrons in the FM will move into the minority band of the SM and lower $\mu_{\rm{FM}}$, until $\mu_{\rm{FM}}=\mu_{\rm min}$ is reached. In this case, the ferromagnet is positively charged, leading to a positive voltage $V_{\rm{S}}=(-\Delta \mu/2)/(-e) = \Delta \mu/(2e) > 0$. All possible combinations of $\bm{S}$ and $\bm{M}_{\rm{FM}}$, as well as the resulting $V_{\bm{S}}$, are illustrated in Figs. 1(c) and 1(d).

These considerations lead to the following conclusion:
\begin{equation}
\begin{cases}
(\bm{M}_{\rm{FM}}\parallel\bm{S})\ \textrm{or}\ (\bm{M}_{\rm{FM}}\parallel\bm{-M}_{\rm{SM}})\ \Leftrightarrow\ V_{\bm{S}}>0\\
(\bm{M}_{\rm{FM}}\parallel\bm{-S})\ \textrm{or}\ (\bm{M}_{\rm{FM}}\parallel\bm{M}_{\rm{SM}})\ \Leftrightarrow\ V_{\bm{S}}<0
\end{cases}.
\end{equation}
This conclusion supports the arguments in Refs. \cite{ele-dection-Dankert, ele-dection-Tang, ele-dection-Lee, ele-dection-Liu, ele-dection-Tian}. It also means that the spin voltage observed in Ref. \cite{ele-dection-Li} corresponds to the spin polarization that is opposite to what the authors of Ref. \cite{ele-dection-Li} thought to be there.

\subsection{Charge-current-induced spin polarization in the topological surface states}

\begin{figure}
\includegraphics[width=0.8 \linewidth]{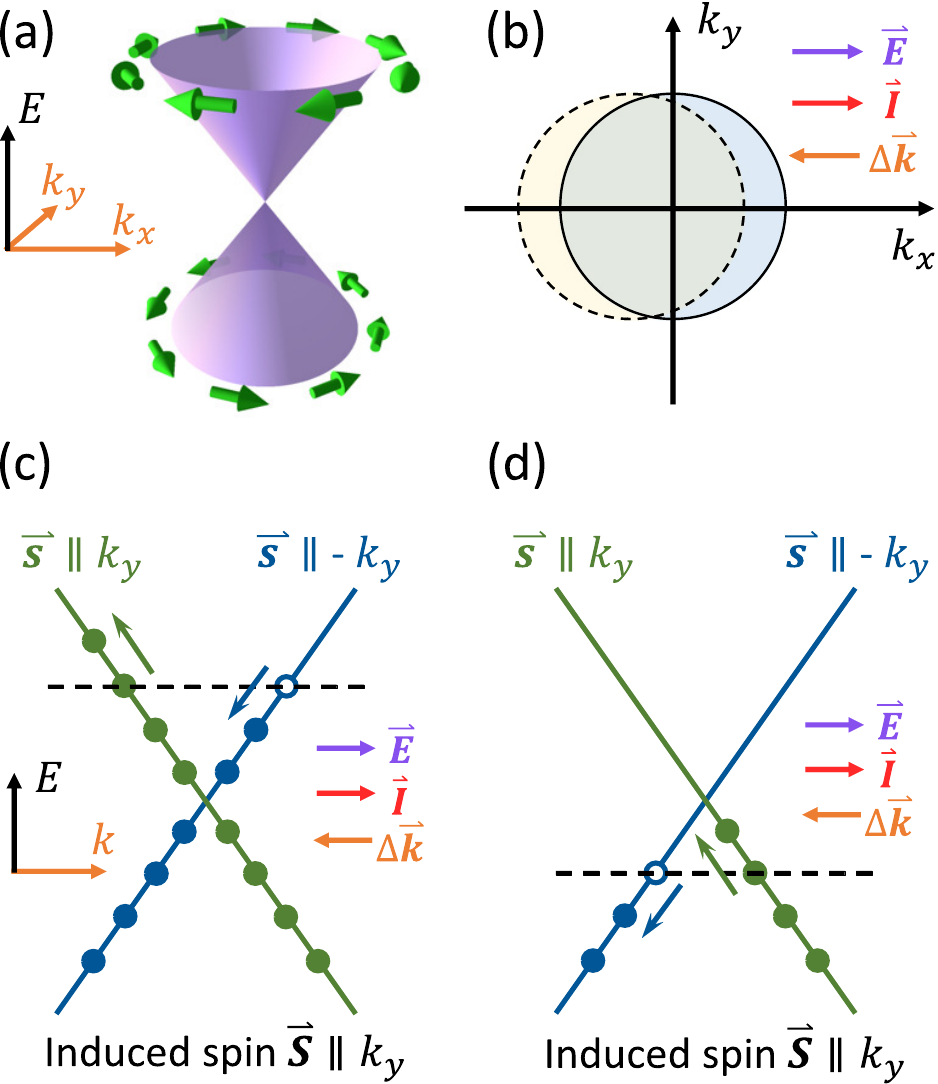}
\caption{\label{fig:Fig2} {
(a) Illustration of the spin textures of the topological surface states. The spin textures above and below the Dirac Point have opposite helicities. (b) Applied electric field along $k_{x}$ generates a charge current in the $k_{x}$ direction, shifting the Fermi surface in the $-k_{x}$ direction due to the negative charge. (c)-(d) Occupations of the Dirac cone states in the $n$- and $p$-type regimes under the influence of charge currents; irrespective of the carrier type, a charge current along $k_{x}$ always makes the $\bm{s}\parallel k_{y}$ branch more populated, resulting in a current-induced spin polarization $\bm{S}$ in the $k_{y}$ direction.
}}
\end{figure}

The TSS of 3D TIs possesses a helical spin texture illustrated in Fig. 2(a); here, the spin vector $\bm{s}$ points to $-\bm{n}\times \bm{v}(\bm{k})$ with $\bm{n}$ the normal vector of the surface and $\bm{v}(\bm{k})$ the group velocity of the Bloch electron with momentum $\bm{k}$. The net spin polarization per unit area, $\bm{S}$, can be written as
\begin{equation}
\bm{S}=\frac{1}{(2\pi)^2}\int_{E(\bm{k})<E_{\rm{F}}}\bm{s}(\bm{k})dk_x dk_y,
\end{equation}
where all the occupied states in the surface Brillouin zone (BZ) contribute to the integral.

In the absence of charge current, the center of the Fermi circle is located at $\bm{k} = 0$ [solid circle in Fig. 2(b)] and the integral in Eq. (2) becomes zero due to symmetry. When the TSS carries a charge current $\bm{I}$ along the $+k_{x}$ direction, the Fermi circle shifts in the $-k_{x}$ direction by $\Delta k$ [dashed circle in Fig. 2(b)]. In this case, the Fermi circle is no longer symmetric with respect to $k_x=0$ and Eq. (2) gives a finite $\bm{S}$ in the $+k_{y}$ direction.

Based on this picture, we now discuss the current-induced spin polarization in $n$-type and $p$-type regions of the Dirac cone. The longitudinal section cut of the Dirac cone along $k_{y}=0$, shown in Figs. 2(c) and 2(d), makes it easy to understand the situation. The spin direction of the green branch is $\bm{s} \parallel k_{y}$ and that of the blue branch is $\bm{s} \parallel -k_{y}$. An electric field $\bm{E}$ in the $+k_x$ direction drives a charge current $\bm{I}$ along the $+k_x$ direction and shifts the Fermi surface in the $-k_x$ direction, as illustrated in Figs. 2(c) and 2(d) for the $n$- and $p$-type regions. Notice that the shift of the Fermi surface always makes the green branch ($\bm{s}\parallel k_{y}$) more populated, no matter if the Fermi level is above or below the Dirac point; this is because the electrons on the green branch are always accelerated by $\bm{E} \parallel k_x$, since their Fermi velocity is along $-k_x$. On the contrary, electrons on the blue branch are always decelerated by $\bm{E} \parallel k_x$. The larger population in the $\bm{s} \parallel k_{y}$ branch gives an induced $\bm{S}$ in the $+k_{y}$ direction. Therefore, for the $\bm{S}$ originating from the TSS, the polarity is always along $\bm{n}\times \bm{I}$ irrespective of the carrier type. At first glance, this conclusion may look counter-intuitive, because the spin helicity reverses when the Fermi level crosses the Dirac point. However, for a fixed $\bm{k}$, the group velocity $\bm{v}(\bm{k})$ also changes sign across the Dirac point. This means that the sign change in $\bm{v}(\bm{k})$ counteracts the spin-helicity reversal and leave the orientation of $\bm{S}$ unchanged.

The above discussion implies that the TSS of 3D TIs can only account for the current-induced $\bm{S}$ directed to $\bm{n} \times \bm{I}$. If the detected $\bm{S}$ is along the $-\bm{n}\times \bm{I}$ direction, one must consider a different origin.


\section{Experiment}

\subsection{Sample preparation and device fabrications}

\begin{figure}
\includegraphics[width=0.9 \linewidth]{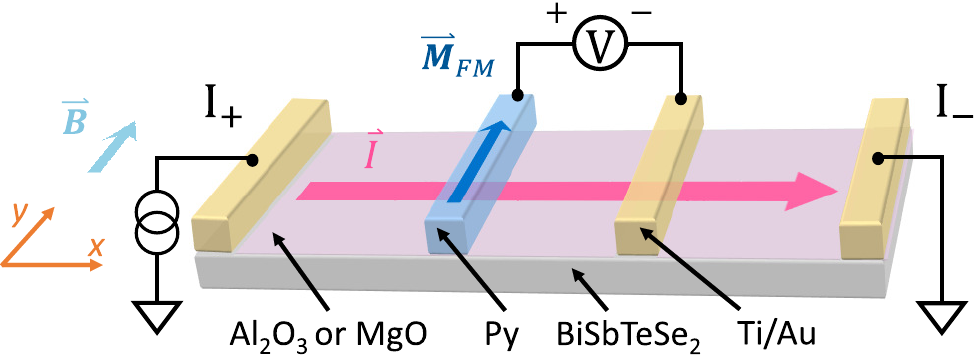}
\caption{\label{fig:Fig3} {
Schematic drawing of the device concept and measurement configuration for spin detection. Several Ti/Au normal-metal contacts and at least one Py ferromagnetic tunneling contact are fabricated on top of a BiSbTeSe$_{2}$ flake. The spin detection is performed in a four-terminal configuration.
}}
\end{figure}

\begin{figure*}
\includegraphics[width=1 \linewidth]{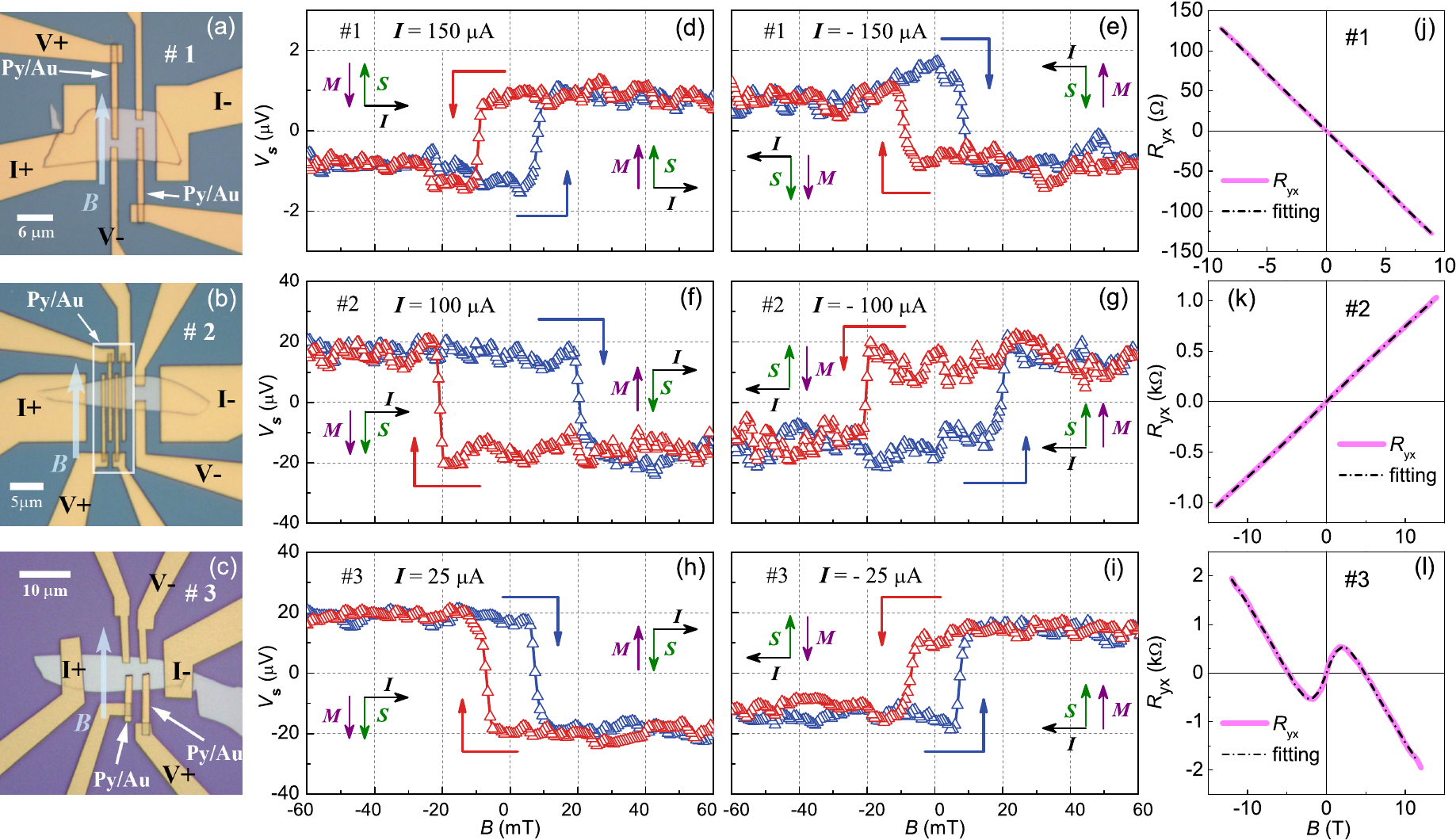}
\caption{\label{fig:Fig4} {
(a, b, c) Optical images of devices \#1, \#2, and \#3, respectively. Measurement configurations are depicted in the pictures. (d)-(i) Spin-dependent voltage $V_{\bm{S}}$ vs. in-plane magnetic field $B$ at both positive and negative bias currents, measured in devices \#1 (d-e), \#2 (f-g), and \#3 (h-i); trivial parabolic background has been subtracted. The polarity of the hysteresis loops of devices \#2 and \#3 is opposite to that of device \#1. (j, k, l) $R_{yx}(B)$ data measured in perpendicular magnetic fields for devices \#1, \#2, and \#3, respectively. For device \#2, the $R_{yx}(B)$ data were measured using the pair of Ti/Au contacts next to the FM contacts; for devices \#1 and \#3, the $R_{yx}(B)$ measurements were performed using the same contact configuration as the spin voltage measurements. The dash-dotted lines are the fits to the data explained in the text. All the data were taken at 4.2 K.
}}
\end{figure*}

BiSbTeSe$_2$ is a bulk-insulating TI with the Fermi level located slightly below the Dirac point \cite{ARPES-BSTS2-Arakane}. We have investigated the current-induced spin polarization in various devices based on exfoliated flakes of BiSbTeSe$_2$. The results presented in this paper were obtained in three typical devices, labeled \#1, \#2 and \#3.

High quality BiSbTeSe$_2$ single crystals were grown by a modified Bridgman method. Thin BiSbTeSe$_2$ flakes were exfoliated from single crystals and transferred onto Si/SiO$_{2}$ substrates. The flakes were examined under a laser confocal microscope. Those with uniform thickness and flat surfaces were selected for device fabrications. The thicknesses of the flakes used in devices \#1, \#2, and \#3 were 172, 82, and 54 nm, respectively.

After the exfoliation, an Al$_{2}$O$_{3}$ or MgO tunnel barrier was deposited on top of the flakes. Devices \#1 and \#2 had the Al$_{2}$O$_{3}$ barrier, which was fabricated by first evaporating 0.7-nm-thick Al in a thermal evaporator and then \textit{in-situ} oxidizing the Al layer with pure oxygen. The MgO barrier was used in device \#3, for which about 2-nm-thick MgO was directly deposited from a MgO source by electron-beam (EB) evaporation. The tunnel barrier proved to be very important for the spin detection. No spin signal was observed in devices with too thin or too thick a tunnel barrier.

The FM and normal-metal contacts were fabricated in two subsequent EB lithography steps. The FM electrodes are made of 30-nm-thick Ni$_{0.81}$Fe$_{0.19}$ (Py) capped with about 160-nm-thick Au. The Au layer prevents Py from oxidation and ensures a reliable electrical connection over the edges of the flakes. The normal-metal electrodes are made of Ti/Au, with the thicknesses of Ti and Au layers being 3 and 190 nm, respectively. Before the deposition of the Ti/Au electrodes, the tunnel barrier in the contact area defined by EB lithography had been removed by shortly dipping the substrate into a diluted tetramethyl-ammonium hydroxide solution. The device concept is schematically illustrated in Fig. 3, and the optical photographs of the devices are shown in Figs. 4(a)-4(c).

\subsection{Chemical potential tuning and characterization}

As-grown BiSbTeSe$_2$ crystals are $p$-type, with the bulk carrier density down to $\sim10^{16}$ $\rm{cm^{-3}}$ \cite{BSTS2-Ren}. According to our experience, if high-temperature baking is avoided during the lithography process, the carrier density in BiSbTeSe$_2$ flakes can be kept low even after they are fabricated into nano-devices; on the other hand, once the flakes are baked at $T >$ 150$^{\circ}$C for several minutes, they will be strongly $n$-type doped and the Fermi level will move into the bulk conduction band. Based on this knowledge, we have managed to obtain BiSbTeSe$_2$ devices with various carrier types and concentrations. For example, during the device fabrication process, device \#1 was baked at $T$ = 170$^{\circ}$C twice to cure the ZEP520A resist, each time for 3 minutes; in contrast, for devices \#2 and \#3, the curing of resist was done at $T$ = 110$^{\circ}$C for 20 minutes, so that the flakes did not experience any high-temperature baking.

The difference in baking temperature resulted in different carrier concentrations. To infer the carrier type and density, we measured the Hall resistivity $R_{yx}$ of all the devices at $T=4.2$ K, as plotted in Figs. 4(j)-4(l).
Device \#1 showed a linear $R_{yx}(B)$ behavior with a negative slope [Fig. 4(j)]; using the formula $R_{yx}=(1/n_{\rm{2d}}e)B$, we obtain the two-dimensional (2D) carrier density $n_{\rm{2d}}=-4.4 \times 10^{13}$ cm$^{-2}$. Here, the negative sign means $n$-type carriers.
The $R_{yx}(B)$ behavior of device \#2 was also linear, but it has a positive slope much larger in absolute value [Fig. 4(k)], indicating $p$-type carriers with much lower carrier density; the linear fit gives $n_{\rm{2d}}=8.5 \times 10^{12}$ cm$^{-2}$.
A highly nonlinear $R_{yx}(B)$ curve was obtained in device \#3 [Fig. 4(l)], indicating the coexistence of $n$-type and $p$-type channels. To obtain the carrier densities of both channels, we fit the $R_{yx}(B)$ curve to the expression given by the two-band model \cite{TI-review-Ando, topgate-Yang},
$$
R_{yx}=\left(\frac{B}{e}\right)\frac{\left(n_1\mu_1^2+n_2\mu_2^2\right)+B^2\mu_1^2\mu_2^2\left(n_1+n_2\right)}{\left(\vert n_1\vert\mu_1+\vert n_2\vert\mu_2\right)^2+B^2\mu_1^2\mu_2^2\left(n_1+n_2\right)^2},
$$
where $n_1$, $\mu_1$, $n_2$, and $\mu_2$ are the 2D carrier density and the mobility of the 1st and 2nd channels, respectively. Those parameters are constrained by the sheet resistance in zero field which is expressed as
$$
\left.R_{\rm sh}\right|_{B=0}=\frac{1}{e\left(\vert n_1\vert\mu_1+\vert n_2\vert\mu_2\right)}.
$$
The fitting gives $n_1 = -3.1 \times 10^{12}$ cm$^{-2}$, $n_2 = 6.8 \times 10^{10}$ cm$^{-2}$, $\mu_1$ = 275 cm$^2$/Vs and $\mu_2$ = 4068 cm$^2$/Vs. This result suggests that the top and bottom surfaces have different types of carriers, but they both have chemical potential located close to the Dirac point.

\subsection{Detection of the spin polarization}

The detection of current-induced spin polarization was performed in a four-terminal configuration, as schematically depicted in Fig. 3. A $dc$ current $\bm{I}$ flows along the $x$ direction between the outer Ti/Au contacts. In-plane magnetic field $\bm{B}$ is applied in the $y$ direction, which is perpendicular to $\bm{I}$ and parallel to the easy axis of the Py spin detector, to control the magnetization $\bm{M}_{\rm FM}$ in Py. Upon scanning $B$ from negative to positive (and {\it vice versa}), the voltage $V_{\rm{FM}}$ between a Py tunneling contact and a Ti/Au contact is measured as a function of $B$. To reduce noise, the data for each curve were averaged over tens of independent scans.

The measured voltage can be written as $V_{\rm{FM}}=V_{\bm{S}}+V_{0}$, where $V_{\bm{S}}$ is the spin-dependent voltage we are interested in, and $V_{0}$ is a trivial parabolic background mainly contributed by the magnetoresistance of BiSbTeSe$_2$. When the Py magnetization $\bm{M}_{\rm FM}$ switches, $V_{\bm{S}}$ suddenly changes sign, whereas $V_{0}$ does not. Therefore, it is easy to separate $V_{\bm{S}}$ from the measured $V_{\rm{FM}}$.

The $V_{\bm{S}}(B)$ data of all the devices are plotted in Figs. 4(d)-4(i). Curves measured in forward and backward $B$-scans form a hysteresis loop. The half width of the loop corresponds to the coercive field of the Py spin detector, which is 10, 20, and 7 mT for devices \#1, \#2, and \#3, respectively. The coercive field of a nanoscale ferromagnet depends on its shape, and narrower width gives higher coercive field. The width of the Py electrodes in devices \#1, \#2 and \#3 was 1, 0.5, and 1.5 $\rm{\mu m}$, respectively, which is consistent with the observed difference in the coercive field.

\begin{figure}
\includegraphics[width=1 \linewidth]{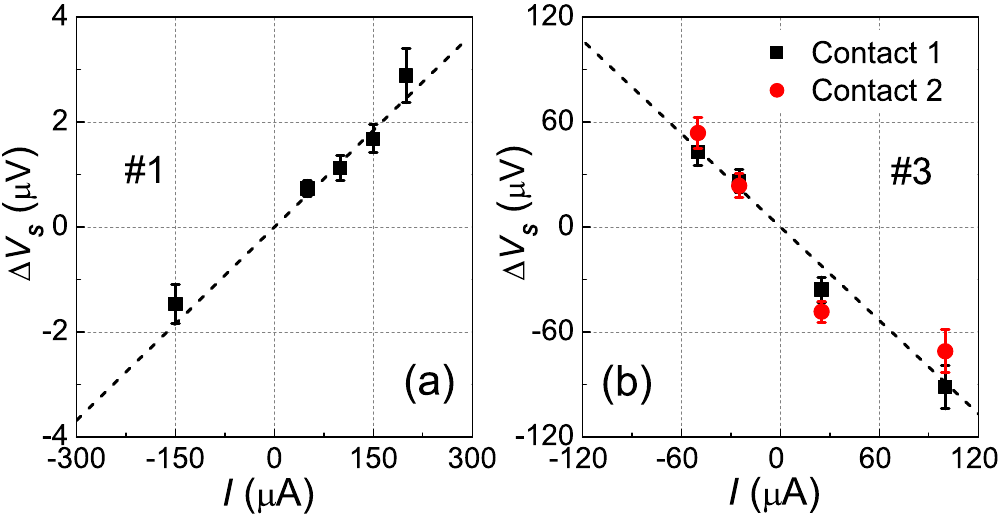}
\caption{\label{fig:Fig5} {
Spin-dependent voltage $\Delta V_{\bm{S}} \equiv V_{\bm{S}}(+B)-V_{\bm{S}}(-B)$ as a function of bias currents, measured in devices \#1 (a) and \#3 (b). The dashed lines are guide to the eyes.
}}
\end{figure}

The amplitude of the spin-dependent voltage can be defined as $\Delta V_{\bm{S}} = V_{\bm{S}}(+B)-V_{\bm{S}}(-B)$. With this definition, the sign of $\Delta V_{\bm{S}}$ is determined by the direction of the jump in the $V_{\bm{S}}(B)$ loop, and hence implies the orientation of the detected $\bm{S}$. For instance, $\Delta V_{\bm{S}}>0$ is obtained when $V_{\bm{S}}>0$ shows up for $B>0$ (i.e. $\bm{B} \parallel y$), and this is the expected sign of $\Delta V_{\bm{S}}$ for $\bm{S} \parallel \bm{B}$ [see Eq. (1)] and thus we can conclude $\bm{S}\parallel y$. Similarly, $\Delta V_{\bm{S}}<0$ implies $\bm{S}\parallel -y$. Hence, the following relation exists between $\Delta V_{\bm{S}}$ and $\bm{S}$:
\begin{equation*}
\begin{cases}
\Delta V_{\bm{S}}>0 \,\Leftrightarrow \, \bm{S}\parallel y\\
\Delta V_{\bm{S}}<0 \,\Leftrightarrow \, \bm{S}\parallel -y
\end{cases}.
\end{equation*}

\begin{figure}
\includegraphics[width=0.75 \linewidth]{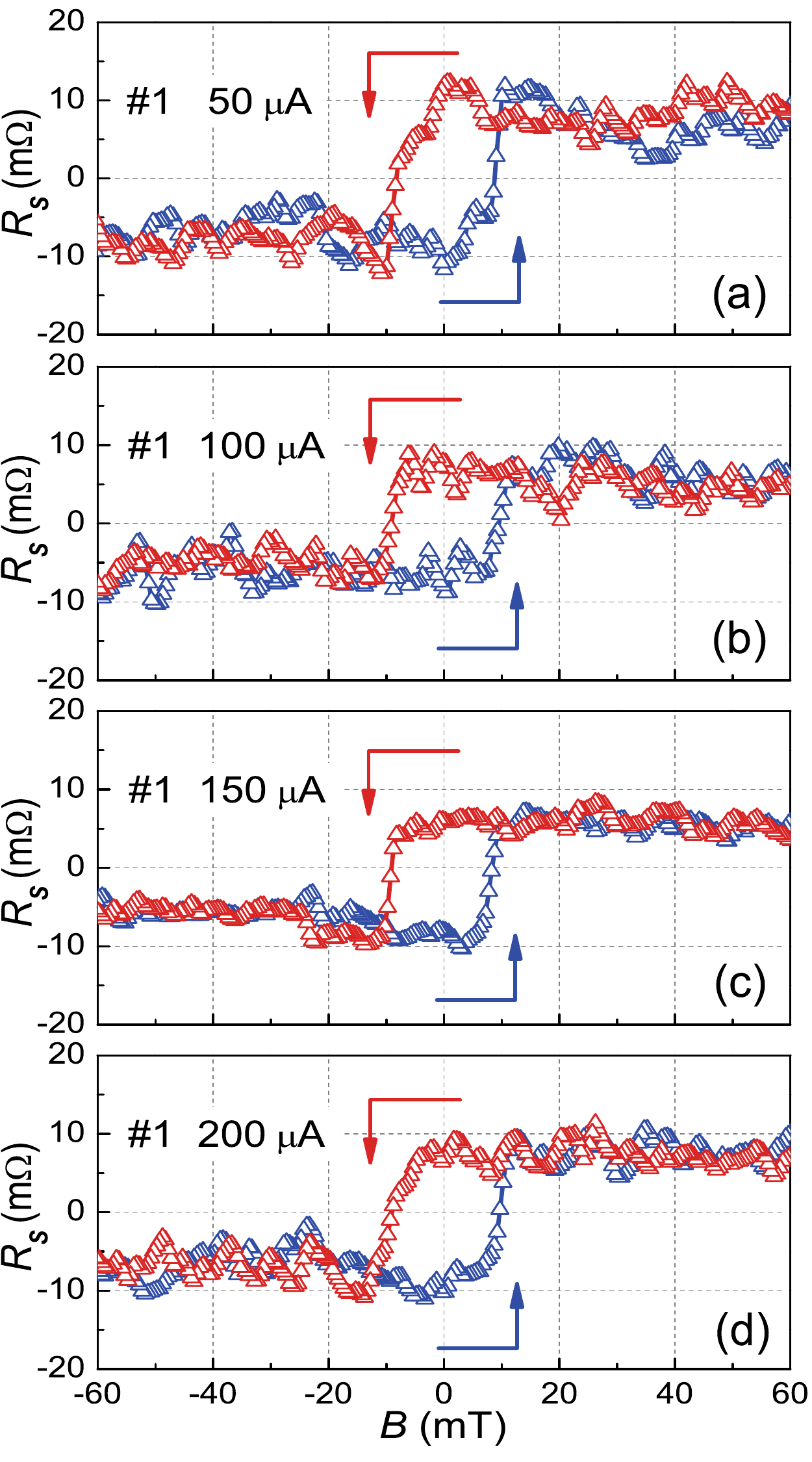}
\caption{\label{fig:Fig6} {
Spin resistance $R_{s} \equiv V_{\bm{S}}/I$ vs. $B$ measured in device \#1 with various bias currents [50, 100, 150, and 200 $\mu$A shown in panels (a)--(d)], demonstrating the ohmicity.
}}
\end{figure}

In our experiment, $V_{\bm{S}}(B)$ loops were measured at both positive and negative currents for each device. The reversal in $\bm{I}$ resulted in the sign change of $\Delta V_{\bm{S}}$, indicating that the orientation of $\bm{S}$ is controlled by the direction of $\bm{I}$. This is consistent with the existence of a helical spin texture in the current-carrying states. However, as one can see in Figs. 4(d), 4(f), and 4(h), the orientation of the current-induced $\bm{S}$ is apparently different between device \#1 and the others. To be specific, our data imply $\bm{S}\parallel y$ for a positive $I$ in device \#1, while it is the opposite in devices \#2 and \#3. This means that the underlying spin polarization is $\bm{S} \parallel (\bm{n}\times \bm{I})$ in the former and $\bm{S} \parallel -(\bm{n}\times \bm{I})$ in the latter. According to the discussion in Sec. II, only the former can be attributed to the TSS of BiSbTeSe$_2$. The main result of the present experiment is the reproducible observation of the opposite polarity, which points to the possibility of utilizing the reversed spin helicity in TI-based devices.

In passing, a linear dependence between $\Delta V_{\bm{S}}$ and $I$ was observed in the $V_{\bm{S}}(B)$ loops measured at various bias currents, as shown in Fig. 5 for devices \#1 and \#3. This linear dependence allows us to define a current-independent spin resistance $R_{\bm{S}} \equiv V_{\bm{S}}/I$. As an example, $R_{\bm{S}}(B)$ data for various $I$ values are shown in Fig. 6 for device \#1.

\section{Discussions}

\subsection{Possible origins of the unexpected spin signal}

As discussed in Sec. III, the $V_{\bm{S}}(B)$ signal in devices \#2 and \#3 cannot be interpreted to be due to the TSS of a 3D TI. Here, we discuss its possible origins.

One possibility is that the observed $V_{\bm{S}}(B)$ is not stemming from a spin polarization but due to some artifacts like local Hall voltages. It was proposed that Hall voltages can be locally induced by fringe fields of the FM electrodes and could give rise to a signal similar to the $V_{\bm{S}}(B)$ loop \cite{FringeField-Vries}. However, one can rule out this possibility in our experiments for the following reasons. First, if the observed $V_{\bm{S}}(B)$ loops are caused by some trivial reasons like the local Hall effect, the signal should be insensitive to the details of the tunnel barrier, and thus it should be observed in all devices with similar FM contacts. In this regard, the hysteretic signals in our experiment are very sensitive to the quality and thickness of the tunnel barrier; we have investigated more than a dozen devices, and the hysteretic voltages were only observed in those devices where the thickness of the tunnel barrier falls into a narrow window and the resulting contact resistance takes a suitable value (5--50 k$\Omega$). The voltage loops were never observed in devices with too thin or too thick a tunnel barriers. Second, specifically for the local Hall effect, since the direction of the unbalanced fringe field depends on both the local geometry of the FM electrode and the morphology of the TI sample \cite{FringeField-Vries}, the polarities of the observed voltage loops should be \textit{random}. Therefore, the voltage loops observed at different FM contacts made on the same device could present opposite polarities. However, in our experiment the voltage loops measured in the same device consistently showed the same polarity at different FM contacts. As an example, the data measured on two different FM contacts in device \#3 are plotted in Fig. 5(b).

To conclusively rule out the possibility that the observed voltage loops were caused by some artifacts (local Hall effect in the TI ﬂake, anomalous Nernsteffect in the FM lead, adverse effects of side contacts, etc.), we have performed a control experiment on a device which was fabricated in the same way as the other devices except that it does not have a tunnel barrier beneath the FM electrodes. As shown in the Supplemental Material (Fig. S2) \cite{Supplemental}, this control device yielded null result even with a high current of 200 $\mu$A. Since all the spurious origins of the voltage loops so far discussed do not require a tunnel contact, the complete absence of a voltage hysteresis in our control device safely rules out the artifacts as the origin of the observed signal.  

Given that the voltage signal is genuinely of spin polarization origin, the most likely origin of the opposite spin voltages observed in devices \#2 and \#3 is the electronic states which have an opposite spin helicity to that of the TSS. To the best of our knowledge, the only known states with such a feature are the Rashba-split 2D states caused by a surface band bending \cite{Rashba-Bianchi, Rashba-King, Rashba-Zhu, Rashba-Bahramy}. When a downward band-bending occurs at the surface of a 3D TI and confines the bulk states in the potential well, the resulting 2D sub-bands present a Rashba spin splitting and form two concentric Fermi circles with opposite spin helicities \cite{Rashba-Bianchi, Rashba-King, Rashba-Zhu, Rashba-Bahramy}, as illustrated in Fig. 7. It has been elucidated that the helicity of the outer Fermi circle is opposite to that of the TSS \cite{Rashba-Bianchi, Rashba-King}, so the contribution of the Rashba-spit states to the spin signal is also opposite.

It is prudent to mention that the opposite spin voltages observed in devices \#2 and \#3 could be due to a {\it negative} 
spin detection efficiency of the FM contacts; namely, it is possible that the minority spin has a larger density of states at $E_F$ compared to the majority spin in a ferromagnet, and in such a case the spin voltage is reversed from the case depicted in Fig. 1. Also, a reversal in the spin voltage may happen if the tunnel barrier has a spin-selective nature. However, this possibility can be largely ruled out for following reasons: (i) For Py/Al$_2$O$_3$ contacts, it is known that the spin polarization of tunneling conductance (SPTC) is positive (i.e. the spin polarization of tunneling current is parallel to the majority spin in Py) \cite{SignProblem-Tsymbal}. (ii) The Py/Al$_2$O$_3$ contacts in devices \#1 and \#2 were made with essentially the same conditions, so it is unlikely that the SPTC is reversed between the two. (iii) By assuming a positive SPTC, our results are consistent with the majority of existing experiments (Ref. 14-19), and can be naturally explained by the band-bending theory we discuss in detail later; if we assume a negative SPTC, the experimental results are not understandable. Nevertheless, we cannot completely exclude the remote possibility that the Py/Al$_2$O$_3$ contact in device \#2 and the Py/MgO contact in device \#3 somehow had an unusual negative SPTC due to unknown reasons; if it were the case, the present conclusion would need reconsideration.


In passing, very recently Li and Appelbaum \cite{critism-Li} reported an experiment in which they observed a similar voltage loop at a FM tunnel contact fabricated on a Au film, and they claimed that such a result discredits the experiments on TI spin devices. However, since Au films are known to exhibit a giant spin Hall effect \cite{SpinHall-Seki}, when a charge current flows through a Au film, an in-plane spin polarization is naturally generated on both top and bottom surfaces and detected by the FM tunnel contact. Hence, their experiment seems to be a confirmation that Au is not a simple metal but a spin-orbit active material.

\begin{figure}
\includegraphics[width=0.8 \linewidth]{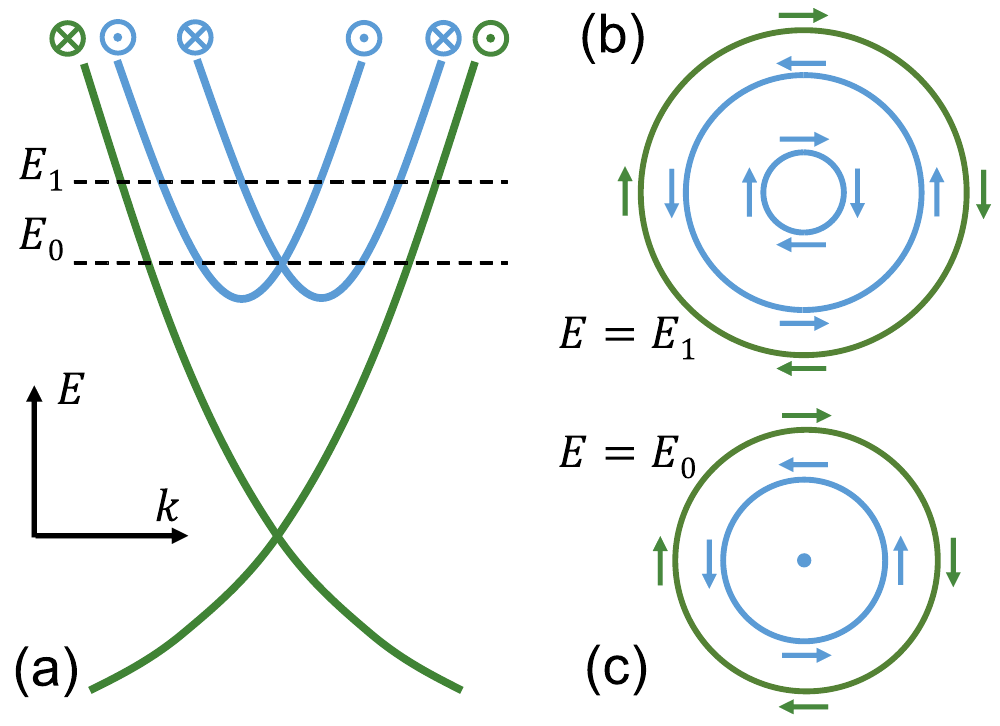}
\caption{\label{fig:Fig7} {
(a) Schematic picture of the Rashba-split 2D states coexisting with the topological surface states. (b)-(c) The constant-energy cuts of the bands at energies $E_1$ and $E_0$ indicated in (a); the Rashba spin-orbit coupling forces the Fermi circle of the 2D states to split into two concentric circles with opposite helicity.
}}
\end{figure}

\subsection{Work function mismatch and band bending}

To judge whether the Rashba-split 2D states are indeed relevant to our devices \#2 and \#3, it is important to understand the band alignment at the Py-TI interface and the work functions of Py and BiSbTeSe$_2$. The work function of Ni$_{0.81}$Fe$_{0.19}$ permalloy is known to be $\Phi_{\rm{Py}}$ = 4.83 eV \cite{WorkFunction-Py-Saito}, which means that the chemical potential of Py lies 4.83 eV below the vacuum level. The work function of our as-grown BiSbTeSe$_2$ crystals was measured by photoemission to be $\Phi_{\rm{BSTS}}$ = 5.20 eV \cite{WorkFunction-BSTS2-Saito}; since the Dirac point of the TSS in as-grown BiSbTeSe$_2$ is located very close to the chemical potential \cite{ARPES-BSTS2-Arakane}, one can see that the Dirac point lies $\sim$5.20 eV below the vacuum level. Therefore, the band alignment of the Py-BiSbTeSe$_2$ interface is such that the Dirac point of the TI side always comes $\sim$0.37 eV below the chemical potential of the metallic Py side.

If the chemical potential of the BiSbTeSe$_2$ flake used in the device is unchanged from the as-grown state, there occurs a charge accumulation in BiSbTeSe$_2$ near the interface to compensate for the $\sim$0.37 eV difference in the chemical potential when the two materials are joined, and this is the basic mechanism of the band bending [compare Figs. 8(c) and 8(d)]. However, when the chemical potential in BiSbTeSe$_2$ is shifted during the device fabrication process, $\Phi_{\rm{BSTS}}$ will change and the strength of the band bending will be different, although the band alignment at the interface is always fixed [compare Figs. 8(b) and 8(d)]. Note that the existence of a tunnel barrier does {\it not} affect the band bending, as long as the barrier is thin enough to allow electrons to flow until the electrochemical potentials of the two sides (TI and Py) equilibrate. The above argument for the band bending may not hold when there are a high density of defect states at the interface, because they would pin the chemical potential; however, our previous top-gating experiments have shown that a thin layer of Al$_2$O$_3$ deposited on a TI surface does not introduce any noticeable defect states at the interface if the deposition is done at temperatures less than 100$^{\circ}$C \cite{topgate-Yang}. In fact, since the (111) surface of tetradymite TI materials does not have dangling bonds, a high density of defect states are not expected at the interface.  

\begin{figure}
\includegraphics[width=1 \linewidth]{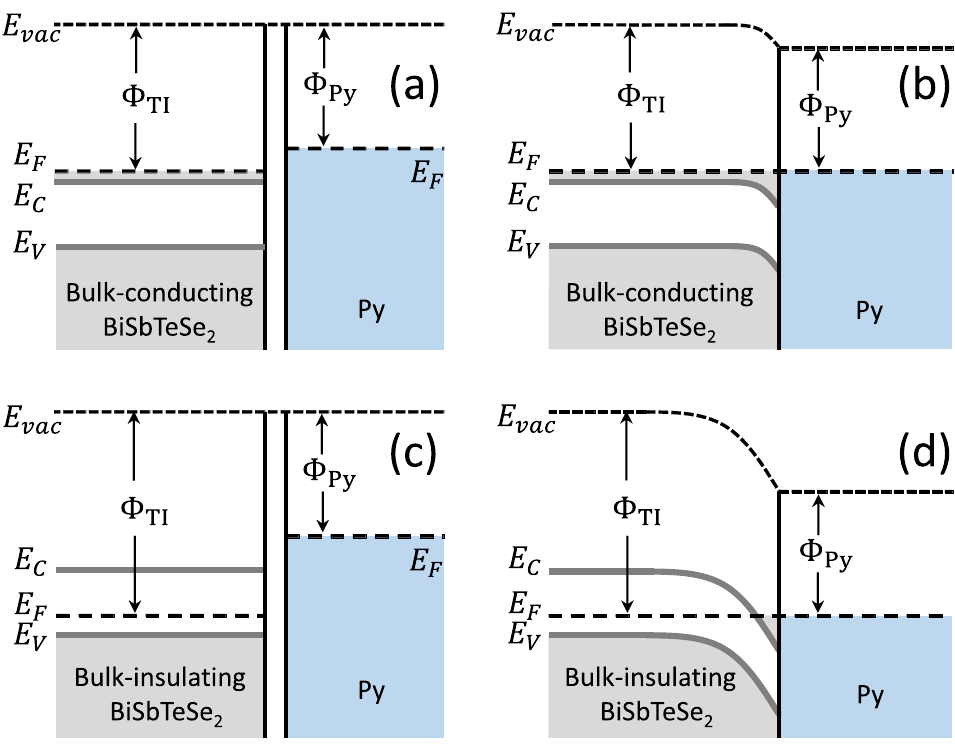}
\caption{\label{fig:Fig8} {
(a)-(b) Energy band diagrams of electron-doped BiSbTeSe$_{2}$ and Py when they are spatially isolated (a) and in contact to form a junction (b); slight band bending occurs at the interface due to the small work function deference. (c)-(d) Similar band diagrams of as-grown BiSbTeSe$_{2}$ and Py. The band bending is more significant because of the larger work function deference.
}}
\end{figure}

We emphasize that the Fermi level of the TI surface beneath the metallic FM electrode is always fixed and will not change with the chemical potential in the TI sample, as long as the electron exchange is allowed through the tunnel barrier. This means that one cannot control the Fermi level of the portions of the TSS at which the current-induced spin polarization is detected by the FM detector, even if the Fermi level of the rest of the TSS can be controlled by electrostatic gating. In other words, even when a sign change in the carrier type is observed upon electrostatic gating in a spin-voltage device (as was the case in Ref. \cite{ele-dection-Lee}), such a change is occurring only in the portion of the sample that is {\it not} in contact with metallic electrodes. Therefore, the gate-voltage-dependences of the spin voltages (such as those reported Ref. \cite{ele-dection-Lee}) are not a result of the gating of the Fermi level of the TSS being probed, but are most likely a reflection of the current redistribution in the TI sample.

\subsection{Fermi level and band diagram}

We have estimated the Fermi level $E_{\rm F}$ in all three devices according to the carrier density obtained from $R_{yx}$ \cite{Supplemental}. Taking the Dirac point as $E=0$, the $E_{\rm F}$ in device \#1 is estimated to be 0.24 eV, corresponding to a work function $\Phi_{\#1}=\Phi_{\rm{BSTS}}-E_{\rm{F}}$ = 4.96 eV. Since the bottom of the conduction band is at $E_{\rm{CB}} \simeq$ 0.21 eV \cite{ARPES-BSTS2-Arakane}, the Fermi level of device \#1 is located inside the conduction band and $E_{\rm{F}}-E_{\rm{CB}} \simeq$ 0.03 eV.

In device \#2, $E_{\rm{F}}$ is estimated to be $-0.07$ eV, corresponding to the work function $\Phi_{\#2}$ = 5.27 eV. The top of the valence band is at $E_{\rm{VB}}=-0.08$ eV \cite{ARPES-BSTS2-Arakane}, so the Fermi level of device \#2 is located inside the bulk band gap and $E_{\rm{F}}-E_{\rm{VB}} \approx 0.01$ eV.

The two surfaces of device \#3 have different types of carriers. The $n$-type surface has $E_{\rm{F}}$ = 0.16 eV and $\Phi_{\#3}^{n}$ = 5.04 eV, while the $p$-type surface has $E_{\rm{F}}=-0.01$ eV and $\Phi_{\#3}^{p}$ = 5.21 eV. Without a gating experiment, it is difficult to tell which surface is the top surface; nevertheless, the $E_{\rm{F}}$ of both surfaces are located inside the bulk band gap.

The band diagrams of BiSbTeSe$_2$/Py contacts are illustrated semi-quantitatively in Fig. 8. Due to the difference in the work function, when Py is in contact with BiSbTeSe$_2$ through a tunnel barrier, electrons in Py will move into BiSbTeSe$_2$ to bias the electrostatic potential of the latter, until the electrochemical potential matches at the interface. In this process, a downward band bending occurs in BiSbTeSe$_2$ and moves the surface Fermi level to $E_{\rm{F}}^{\rm surface}=\Phi_{\rm{BSTS}}-\Phi_{\rm{Py}}=0.37$ eV, which is $\sim$0.16 eV above the bottom of the conduction band. As already mentioned, $E_{\rm{F}}^{\rm surface}$ in the contact area is determined by the band structures of BiSbTeSe$_2$ and Py, and the carrier density in the BiSbTeSe$_2$ flakes has no influence on it.

The strength of the band bending is expressed by $\Delta E=\Phi_{\#i}-\Phi_{\rm{Py}}$ (\#$i$ is the device number). Based on the estimates of $\Phi_{\#i}$ presented above, we obtain $\Delta E$ to be 0.13 eV in device \#1, 0.44 eV in device \#2, and either 0.21 or 0.38 eV in device \#3. The strength of the Rashba spin splitting in the confined 2D states is proportional to the electric field in the $z$ direction, which depends linearly on $\Delta E$.
The experiment and model calculation in Ref. \cite{Rashba-Bianchi} showed that a small $\Delta E$ of 0.13 eV would not cause a measurable Rashba spin splitting in Bi$_2$Se$_3$, and hence one would not expect its contribution to the $V_{\bm{S}}$ measurements in device \#1.

According to a photoemission study \cite{Rashba-King}, the Rashba splitting in Bi$_2$Se$_3$ is barely visible for $\Delta E \simeq$ 0.15 eV and becomes as large as $\Delta k_{\rm{F}} \simeq 0.08 \rm{\AA^{-1}}$ for $\Delta E$ = 0.35 eV, giving the Rashba parameter of $\alpha\sim1.3$ eV$\rm{\AA}$. This suggests that in our devices \#2 and \#3, the band bending is strong enough to cause significant contribution of the Rashba-split 2D states in the $V_{\bm{S}}$ measurements.
Hence, the important question is how the contribution from the Rashba-split states can dominate the observed spin signal.
When one looks at the situation, the Fermi circle of the TSS is always larger than that of the Rashba-split states, so the carrier density of the TSS is also larger; if the mobility is the same, the TSS would carry more current. Also, the contribution from the outer Fermi circle of the Rashba-split states is partially canceled by the inner circle; therefore, the spin signal from the TSS, which experiences no such cancellation, would be stronger.

Nevertheless, there are some factors which speak for the Rashba-split states. First, when $\Delta E$ is large, there appear more than one pairs of Rashba-split states at the Fermi level. For instance, a second pair of Rashba-split states has been observed in photoemission experiments \cite{Rashba-Zhu, Rashba-Bahramy}. In such a situation, the total spin signal contributed by all pairs could exceed that from the TSS. Second, as $k_{F}$ increases, the Fermi circles of the TSS start to be deformed and show hexagonal warping \cite{Rashba-Bahramy}. In a deformed Fermi circle, spin is no long strictly perpendicular to the wavevector. Since the Fermi circle of the TSS is larger than the others and the hexagonal warping grows as $\sim k^3$, it is more prone to the deformation and the in-plane spin component gets weaker. Therefore, the hexagonal warping helps the Rashba-split states to win over the TSS in the contribution to the spin signal. Third, as shown in Fig. 7, the inner circle of the Rashba-split states can shrink to a point when $E_{\rm{F}}=E_{0}$. At this point, theoretical calculation \cite{CIS-model-Hong} shows that the spin signal coming from the Rashba-split states can be stronger than that from the TSS. 

\subsection{Spin-detection efficiency}

According to the theory \cite{CIS-model-Hong}, for the current-induced $\bm{S}$ generated by a single helical channel, we have
\begin{equation}
\Delta R_{{\bm{S}}}=\frac{\Delta V_{\bm{S}}}{I}=\frac{h\pi}{e^2Wk_{\rm{F}}}P_{\rm{FM}}P_{\bm{S}},
\end{equation}
where $k_{\rm{F}}$ is the Fermi wave vector of the helical channel, $W$ is the width of the device, $P_{\rm{FM}}$ is the effective spin polarization of the FM detector, and $P_{\bm{S}}$ is the induced spin polarization per unit current. For a Py spin detector, we can take $P_{\rm{FM}} \approx 0.45$ \cite{ele-dection-Tian}. The value of $P_{\bm{S}}$ is determined by the spin texture of the channel and $P_{\bm{S}}=2/\pi$ is expected for an ideal TSS \cite{CIS-model-Hong}.

In reality, the spin-detection efficiency $\eta$ is not 100\% and Eq. (3) is modified to
\begin{equation}
\Delta R_{{\bm{S}}}=\eta\frac{h\pi}{e^2Wk_{\rm{F}}}P_{\rm{FM}}P_{\bm{S}}.
\end{equation}

As discussed in Sec. III-C, for device \#1, the measured spin signal is mostly due to the TSS even though a major fraction of the current is carried by bulk carriers. In this case, $\eta$ can be estimated by using Eq. (4). Specifically, we have $\Delta R_{{\bm{S}}}= 12.3$ m$\Omega$ and $W=8.5$ $\mu$m for device \#1, and the band alignment at the interface fixes the Fermi level of the TSS beneath the FM electrode at $E_{\rm{F}}^{\rm surface}=0.37$ eV, which corresponds to $k_{\rm{F}}=0.12$ $\rm{\AA^{-1}}$ \cite{Supplemental}. These values lead to $\eta \approx$ 0.54\%. Such a low detection efficiency is probably due to the coexistence of the bulk states. The spin signals in devices \#2 and \#3 are supposedly dominated by the Rashba-split states, but the reason for their dominance is not quantitatively understood. Hence, the spin-detection efficiency in devices \#2 and \#3 cannot be estimated at this stage.

\subsection{Implication for spin transistors}

While the exact mechanism of the sign reversal in $V_{\bm{S}}$ is to be elucidated in future, the discovery that the current-induced spin polarization can be switched by changing the chemical potential of a TI is of significant practical importance, because it allows us to conceive a spin-transistor device, in which the output spin polarization is switched by electrostatic gating. In this regard, it is conceivable that when the TI layer is thin enough, back gating can change the chemical potential of the TI throughout thickness, leading to a change in the band bending at the top surface beneath the FM electrode. Such a tuning of the band bending would allow us to control the dominant spin helicity and the spin-transistor operation can be realized. For the exploitation of this intriguing and useful effect, further studies of its mechanism is strongly called for.

\section{Conclusion}

We discovered that the current-induced spin polarization in BiSbTeSe$_2$ flakes can be reversed depending on the Fermi level. In particular, samples with a small Fermi energy present the spin polarization that is opposite to what is expected for the topological surface states. This is most likely due to the contribution from Rashba-split 2D states created by a strong band bending occurring at the interface of the TI and the ferromagnetic spin detector. While its exact mechanism is to be elucidated in future studies, this effect provides an operation principle for a spin-transistor device in which the output spin polarization is controlled by electrostatic gating.

\begin{acknowledgments}
We thank M. Novak and Z. Wang for their help in crystal growth, Y. Maekawa for her help in device fabrications, and S. Sasaki for his help in various aspects of the experiments. We also thank T. Sato and his group for measuring the work function of BiSbTeSe$_2$. This work was supported by JSPS (KAKENHI 25220708) and DFG (CRC1238 ``Control and Dynamics of Quantum Materials", Projects A04 and C02).
\end{acknowledgments}

\onecolumngrid
\newpage

\renewcommand{\thefigure}{S\arabic{figure}} 
\renewcommand{\thesection}{S\arabic{section}} 

\setcounter{figure}{0}
\setcounter{section}{0}

\begin{flushleft} 
{\Large {\bf Supplemental Material}}
\end{flushleft} 

\section{Estimation of Fermi Levels in B\lowercase{i}S\lowercase{b}T\lowercase{e}S\lowercase{e}$_2$ Flakes}

\begin{figure}[b]
\includegraphics[width=0.7 \linewidth]{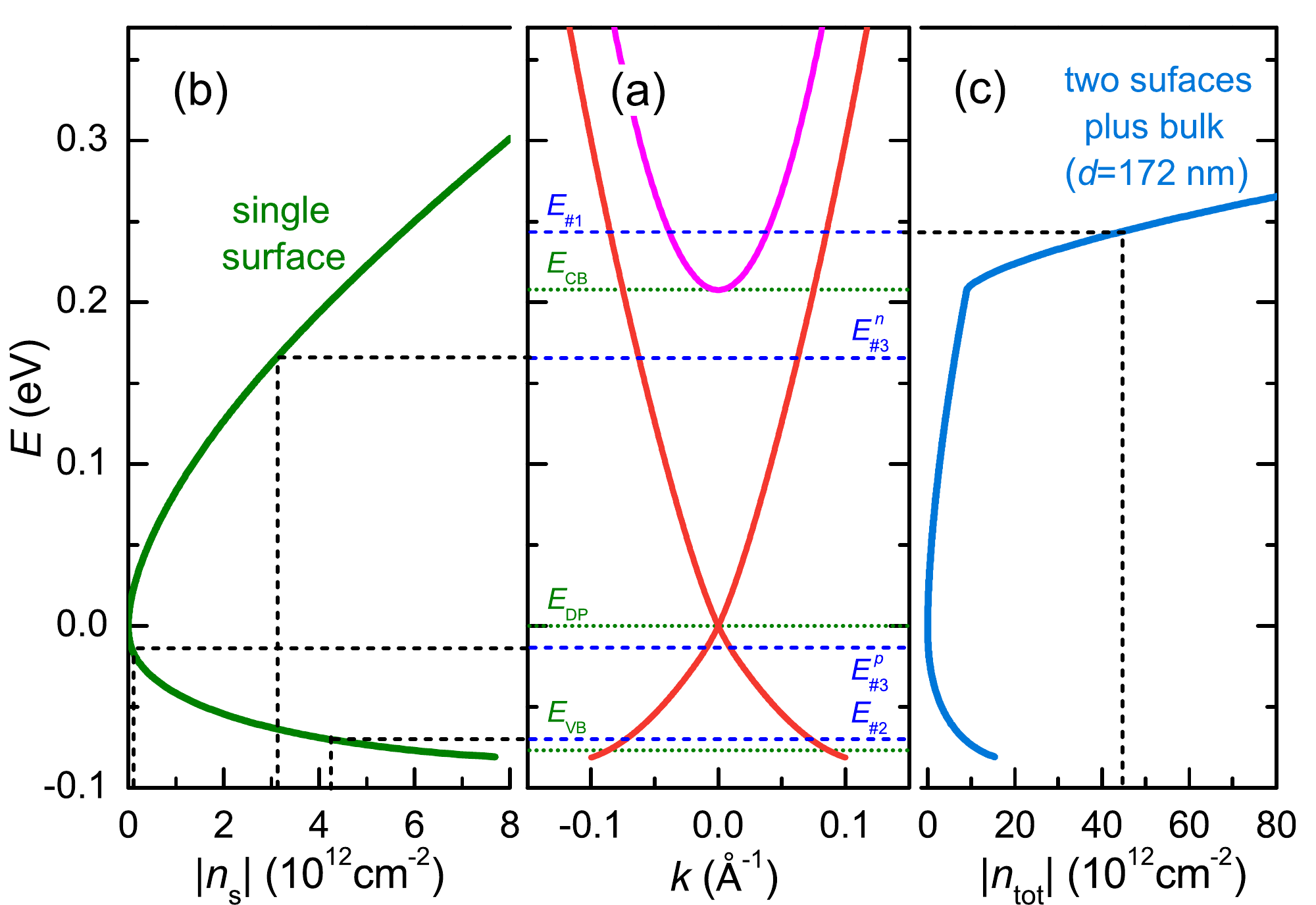}
\caption{\label{fig:FigS1} {
(a) Energy dispersions of bulk states (orange curve) and surface states (red curve) of BiSbTeSe$_{2}$. Positions of the conduction band bottom, Dirac point, and the valence band top are illustrated by the green dotted lines. The estimated Fermi-level positions are indicated by the cyan dashed lines. (b)-(c) Plots of Fermi energy versus carrier density for (b) a single surface and (c) two surfaces plus bulk. The thickness of device \#1 ($d=172$ nm) is used for calculating the curve in (c).
}}
\end{figure}

To estimate the Fermi level from the carrier density, one needs to know the dispersion relations of both the surface states and the bulk states of BiSbTeSe$_{2}$. For this purpose, we took the surface-state dispersion of BiSbTeSe$_{2}$ obtained by angle-resolved photoemission spectroscopy (ARPES) \cite{ARPES-BSTS2-Arakane} and smoothly extrapolated the curve to a wider energy scale, as plotted in Fig. \ref{fig:FigS1}. Also, according to the same ARPES result \cite{ARPES-BSTS2-Arakane}, the conduction band bottom and the valence band top are located at $E_{\rm{CB}} \approx 0.21$ eV and $E_{\rm{VB}} \approx -0.08$ eV, respectively, as indicated by green dotted lines in Fig. \ref{fig:FigS1}(a). Note that $E$ is measured from the Dirac point of the surface state.

As for the bulk conduction band, since its exact dispersion in BiSbTeSe$_{2}$ is not yet known, we used the dispersion in Bi$_{2}$Se$_{3}$ \cite{ARPES-Bi2Se3-Chen} as an approximation to that in BiSbTeSe$_{2}$. The bulk dispersion in Fig. \ref{fig:FigS1}(a) is a reproduction of the data for Bi$_{2}$Se$_{3}$ reported in Ref. \cite{ARPES-Bi2Se3-Chen} with the location of the band bottom adjusted to $E_{\rm{CB}}=0.21$ eV.

At zero temperature, the carrier density contributed by each band is determined by the volume (or area) enclosed by the Fermi surface (or Fermi circle) of that band. For a topological insulator, the two-dimensional (2D) carrier density of the topological surface states and bulk states can be respectively written as
\begin{equation}
n_{\rm{s}}(k_{F})=\frac{1}{(2\pi)^2}\cdot \pi k_{F}^{2}=\frac{k_{F}^{2}}{4\pi}
\end{equation}
and
\begin{equation}
n_{\rm{b}}(k_{F})=2\cdot \frac{1}{(2\pi)^3}\cdot \frac{4}{3}\pi k_{F}^{3} \cdot d=\frac{k_{F}^{3}d}{3\pi^{2}},
\end{equation}
where $k_{F}$ is the Fermi wave vector and $d$ is the sample thickness. The prefactor 2 in Eq. (2) comes from the spin degeneracy of the bulk states.

For a given carrier density, the corresponding energy level $E$ of the highest occupied state can be calculated by using Eqs. (1) and (2). Figure S1(b) shows such a relation upon considering only the surface sates; the Fermi levels of devices \#2 and \#3 can be read off from this figure, since their carrier densities are low enough for the bulk carriers to be absent. On the other hand, for device \#1 we need to take into account the bulk carriers because of its relatively high carrier density; by assuming that the total 2D carrier density $n_{\rm 2d}$ of device \#1 determined from the Hall resistivity (see Sec. III-B of the main text) can be approximated by $n_{\rm 2d} = 2n_{\rm{s}}+n_{\rm{b}}$, the $E$ vs. $n_{\rm{2d}}$ relation is calculated as shown in Fig. \ref{fig:FigS1}(c). The Fermi level of device \#1 can be estimated from this figure.

As a result of the above procedure, the Fermi levels in all three devices are estimated to be $E_{\rm{\#}1}=$ 0.24 eV, $E_{\rm{\#}2} = -0.07$ eV, $E_{\rm{\#}3}^{n}=$ 0.16 eV and $E_{\rm{\#}3}^{p}=-0.01$ eV for to devices \#1, \#2 and the two surfaces of devices \#3, respectively.

\section{Control Experiment: Device Without Tunnel Barrier}

To rule out the possibility that the observed hysteresis loops in devices \#1 -- \#3 are due to artifacts like the local Hall effect (within the TI) or the anomalous Nernst effect (within the FM), we fabricated a device (labeled S1) for a control experiment. The thickness of the flake used in device S1 is 298 nm. The ferromagnetic (FM) contacts are made of Py/Au, without a tunnel barrier beneath it. The measurement was performed at $T = 1.8$ K.

Since the occurrence and detection of both the local Hall effect and the anomalous Nernst effect do not require a tunnel contact, if the signals observed in devices \#1 -- \#3 were due to such artifacts, similar signals should also be observed in device S1. However, as shown in Fig. S2(b), no hysteresis loop is observed at the excitation current of as large as 200 $\mu$A, the maximum current that is used in this experiment. Therefore, possible artifacts due to the local Hall effect or the anomalous Nernst effect can be safely ruled out.

\begin{figure}
\includegraphics[width=0.75 \linewidth]{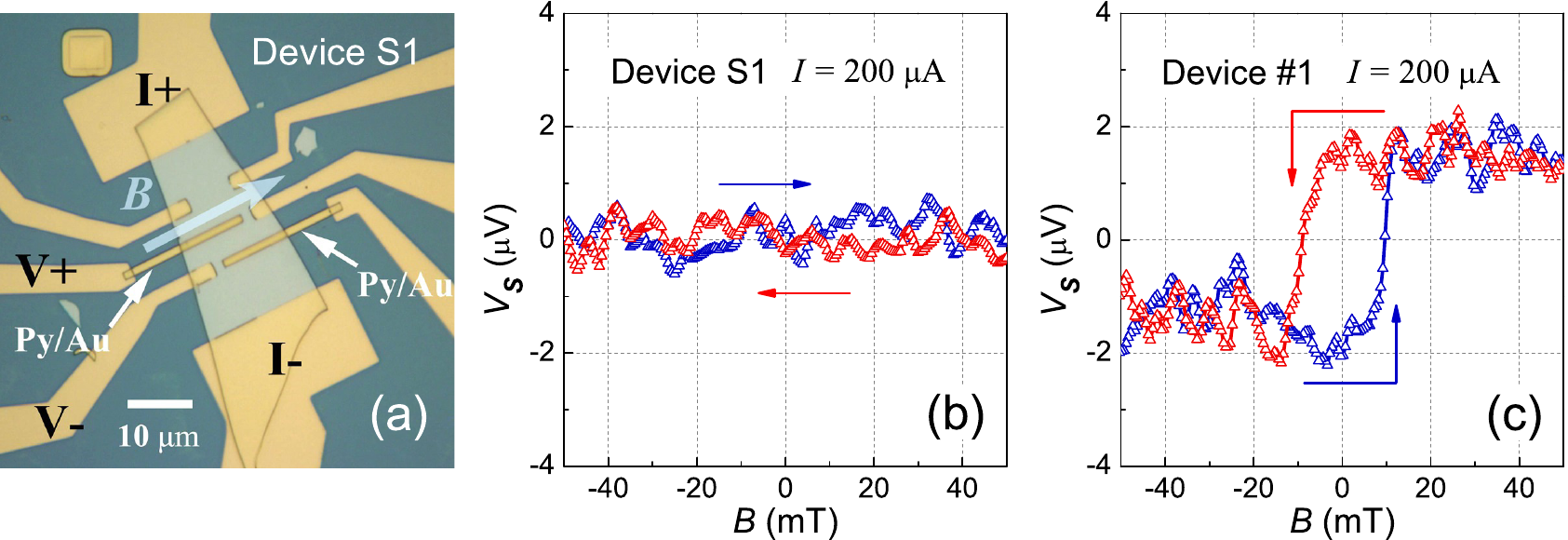}
\caption{\label{fig:FigS2} {
(a) Optical image of device S1. Measurement configuration is depicted in the picture. (b) The $V_{\bm{S}}(B)$ data measured in device S1, showing no hysteresis loop. (c) The $V_{\bm{S}}(B)$ data measured in device \#1, for comparison. The same data are plotted in Fig. 6(d) of the main text.
}}
\end{figure}

\section{Surface Morphology of Exfoliated B\lowercase{i}S\lowercase{b}T\lowercase{e}S\lowercase{e}$_2$ Flakes}

The topological insulator thin films grown by the molecular beam epitaxy (MBE) method usually have triangular terraces on the surface. The fringe fields generated at the edge of such triangular features were proposed to be the possible origin of the local Hall effect \cite{LocalHall-Vries}. Here we show that, in contrast to the MBE-grown thin films, the surfaces of exfoliated BiSbTeSe$_{2}$ flakes are usually very smooth. One can easily find flakes with atomically flat surfaces. In Figs. 3(b)-3(c) we show the atomic-force-microscope (AFM) data measured on the top surface of the flake shown in Fig. 3(a). The surface is atomically flat with a height variation of only $\pm0.25$ nm. It is possible that the morphology difference between MBE-grown thin films and exfoliated flakes is the reason why the local Hall effect reported in Ref. [3] is not observed in our experiment.

\begin{figure}
\includegraphics[width=0.7 \linewidth]{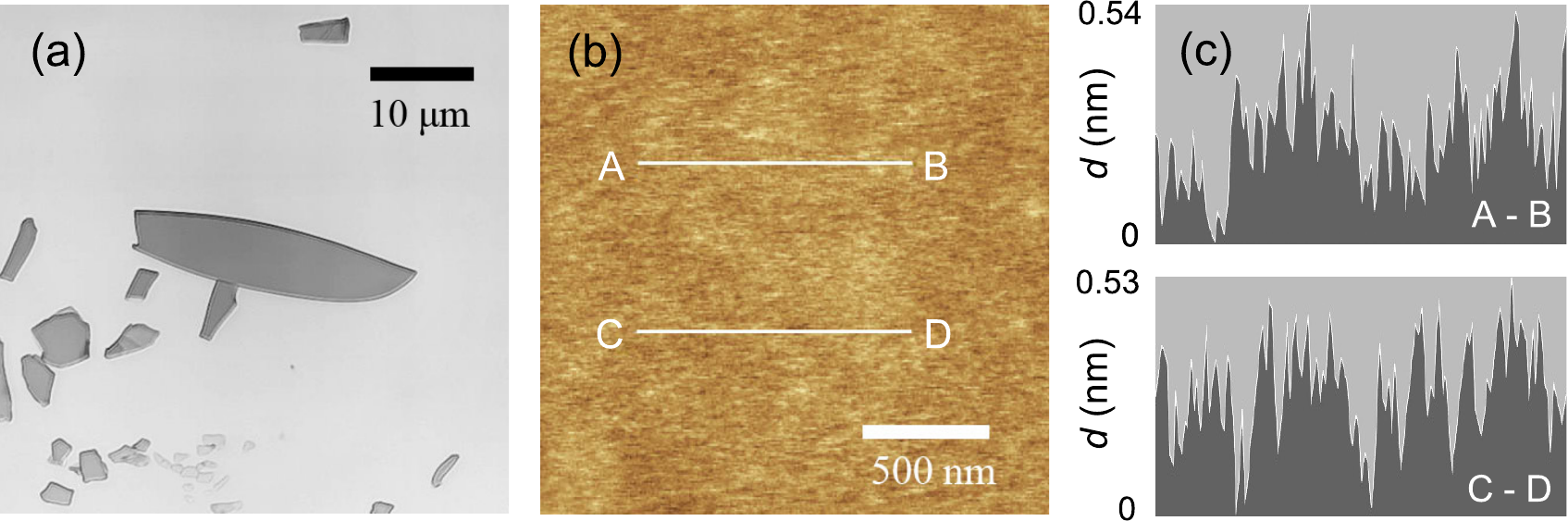}
\caption{\label{fig:FigS3} {
(a) Laser-microscope image of an exfoliated BiSbTeSe$_{2}$ flake. (b) Magnified AFM image of the flake shown in panel (a). (c) Height-profile data along lines A-B and C-D indicated in panel (b).
}}
\end{figure}

\section{$R(T)$ Curves of the Exfoliated B\lowercase{i}S\lowercase{b}T\lowercase{e}S\lowercase{e}$_2$ Flakes}

The $R_{xx}(T)$ curves of two BiSbTeSe$_{2}$ flakes (labeled F1 and F2) are shown in Fig. S4(a). Both flakes F1 and F2 show curved Hall resistance, as plotted in Fig. S4(b). The fitting of the F1 data to the two-band model gives $n_{1}^{F1}=-1.6\times10^{11}$ cm$^{-2}$ and $n_{2}^{F1}=1.8\times10^{12}$ cm$^{-2}$ for the two types of carriers (most likely those on top and bottom surfaces), while the same analysis for the F2 data gives $n_{1}^{F2}=1.3\times10^{11}$ cm$^{-2}$ and $n_{2}^{F2}=1.9\times10^{13}$ cm$^{-2}$.

\begin{figure}
\includegraphics[width=0.8 \linewidth]{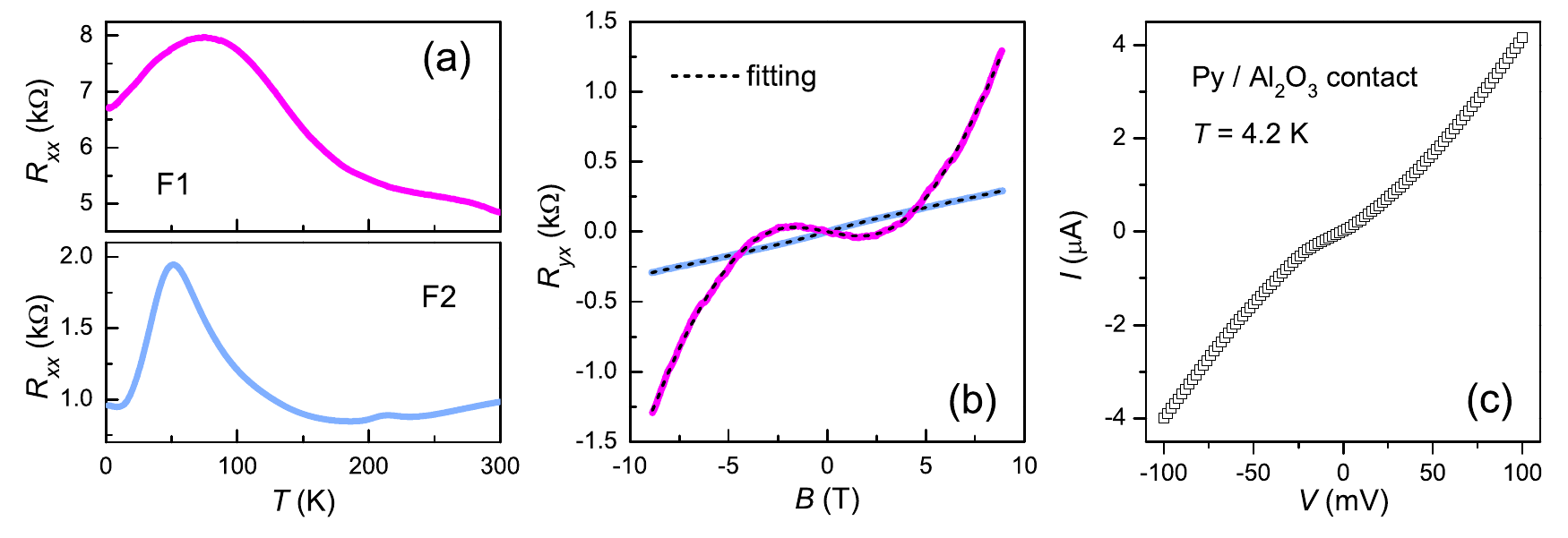}
\caption{\label{fig:FigS4} {
(a) $R_{xx}(T)$ curves of flakes F1 and F2. (b) $R_{yx}(B)$ data of flakes F1 (red) and F2 (blue). Dashed lines are fits to the two-band model. (c) $I$-$V$ curve of an Al$_{2}$O$_{3}$/Py contact made on a BiSbTeSe$_{2}$ flake.
}}
\end{figure}

\section{Additional Information about FM Tunnel Contacts}

The areas of the FM tunnel contacts were $\sim3$ $\mu\rm{m}^{2}$ for device \#2 and $\sim6$ $\mu\rm{m}^{2}$ for devices \#1 and \#3. The resistance-area product of the FM contacts is in the range of $30-200$ $\rm{k}\Omega\, \mu m^{2}$. The current-voltage ($I$-$V$) characteristics of the FM contacts are usually non-linear, suggesting tunneling behavior. The $I$-$V$ curve of a typical Al$_{2}$O$_{3}$/Py contact is shown in Fig. S4(c).


\begin{thebibliography}{10}

\bibitem{TI-review-Hasan} M. Z. Hasan and C. L. Kane, \textit{Colloquium: Topological Insulators}, Rev. Mod. Phys. \textbf{82}, 3045 (2010).

\bibitem{TI-review-Qi} X.-L. Qi and S.-C. Zhang, \textit{Topological Insulators and Superconductors}, Rev. Mod. Phys. \textbf{83}, 1057 (2011).

\bibitem{TI-review-Ando} Y. Ando, \textit{Topological Insulator Materials}, J. Phys. Soc. Jpn. \textbf{82}, 102001 (2013).

\bibitem{TS-theory-Burkov} A. A. Burkov and D. G. Hawthorn, \textit{Spin and Charge Transport on the Surface of a Topological Insulator}, Phys. Rev. Lett. \textbf{105}, 066802 (2010).

\bibitem{TS-theory-Pesin} D. Pesin and A. H. MacDonald, \textit{Spintronics and Pseudospintronics in Graphene and Topological Insulators}, Nat. Mater. \textbf{11}, 409 (2012).

\bibitem{TS-theory-Mahfouzi} F. Mahfouzi, N. Nagaosa, and B. K. Nikoli\'{c}, \textit{Spin-Orbit Coupling Induced Spin-Transfer Torque and Current Polarization in Topological-Insulator/Ferromagnet Vertical Heterostructures}, Phys. Rev. Lett. \textbf{109}, 166602 (2012).

\bibitem{spin-torque-Mellnik} A. R. Mellnik, J. S. Lee, A. Richardella, J. L. Grab, P. J. Mintun, M. H. Fischer, A. Vaezi, A. Manchon, E.-A. Kim, N. Samarth, and D. C. Ralph, \textit{Spin-transfer Torque Generated by a Topological Insulator}, Nature (London) \textbf{511}, 449 (2014).

\bibitem{spin-torque-Fan} Y. Fan, P. Upadhyaya, X. Kou, M. Lang, S. Takei, Z. Wang, J. Tang, L. He, L.-T. Chang, M. Montazeri, G. Yu, W. Jiang, T. Nie, R. N. Schwartz, Y. Tserkovnyak, and K. L. Wang, \textit{Magnetization Switching through Giant Spin-Orbit Torque in a Magnetically Doped Topological Insulator Heterostructure}, Nat. Mater. \textbf{13}, 699 (2014).

\bibitem{spin-torque-Wang} Y. Wang, P. Deorani, K. Banerjee, N. Koirala, M. Brahlek, S. Oh, and H. Yang,
\textit{Topological Surface States Originated Spin-Orbit Torques in $Bi_{2}Se_{3}$}, Phys. Rev. Lett. \textbf{114}, 257202 (2015).

\bibitem{spin-pumping-Shiomi} Y. Shiomi, K. Nomura, Y. Kajiwara, K. Eto, M. Novak, K. Segawa, Y. Ando, and E. Saitoh, \textit{Spin-Electricity Conversion Induced by Spin Injection into Topological Insulators}, Phys. Rev. Lett. \textbf{113}, 196601 (2014).

\bibitem{spin-pumping-Jamali} M. Jamali, J. S. Lee, J. S. Jeong, F. Mahfouzi, Y. Lv, Z. Zhao, B. K. Nikoli\'{c}, K. A. Mkhoyan, N. Samarth, and J.-P. Wang, \textit{Giant Spin Pumping and Inverse Spin Hall Effect in the Presence of Surface and Bulk Spin-Orbit Coupling of Topological Insulator $Bi_{2}Se_{3}$}, Nano Lett. \textbf{15}, 7126 (2015).

\bibitem{spin-pumping-Deorani} P. Deorani, J. Son, K. Banerjee, N. Koirala, M. Brahlek, S. Oh, and H. Yang, \textit{Observation of Inverse Spin Hall Effect in Bismuth Selenide}, Phys. Rev. B \textbf{90}, 094403 (2014).

\bibitem{ele-dection-Li} C. H. Li, O. M. J. vant Erve, J. T. Robinson, Y. Liu, L. Li, and B. T. Jonker, \textit{Electrical Detection of Charge-Current-Induced Spin Polarization Due to Spin-Momentum Locking in $Bi_{2}Se_{3}$}, Nat. Nanotechnol. \textbf{9}, 218 (2014).

\bibitem{ele-dection-Ando} Y. Ando, T. Hamasaki, T. Kurokawa, K. Ichiba, F. Yang, M. Novak, S. Sasaki, K. Segawa, Y. Ando, and M. Shiraishi, \textit{Electrical Detection of the Spin Polarization Due to Charge Flow in the Surface State of the Topological Insulator $Bi_{1.5}Sb_{0.5}Te_{1.7}Se_{1.3}$}, Nano Lett. \textbf{14}, 6226 (2014).

\bibitem{ele-dection-Dankert} A. Dankert, J. Geurs, M. V. Kamalakar, S. Charpentier, and S. P. Dash, \textit{Room Temperature Electrical Detection of Spin Polarized Currents in Topological Insulators}, Nano Lett. \textbf{15}, 7976 (2015).

\bibitem{ele-dection-Tang} J. Tang, L.-T. Chang, X. Kou, K. Murata, E. S. Choi, M. Lang, Y. Fan, Y. Jiang, M. Montazeri, W. Jiang, Y. Wang, L. He, and K. L. Wang, \textit{Electrical Detection of Spin-Polarized Surface States Conduction in $(Bi_{0.53}Sb_{0.47})_{2}Te_{3}$ Topological Insulator}, Nano Lett. \textbf{14}, 5423 (2014).

\bibitem{ele-dection-Liu} L. Liu, A. Richardella, I. Garate, Y. Zhu, N. Samarth, and C.-T. Chen, \textit{Spin-Polarized Tunneling Study of Spin-Momentum Locking in Topological Insulators}, Phys. Rev. B \textbf{91}, 235437 (2015).

\bibitem{ele-dection-Lee} J. S. Lee, A. Richardella, D. R. Hickey, K. A. Mkhoyan, and N. Samarth, \textit{Mapping the Chemical Potential Dependence of Current-Induced Spin Polarization in a Topological Insulator}, Phys. Rev. B \textbf{92}, 155312 (2015).

\bibitem{ele-dection-Tian} J. Tian, I. Miotkowski, S. Hong, and Y. P. Chen, \textit{Electrical Injection and Detection of Spin-Polarized Currents in Topological Insulator $Bi_{2}Te_{2}Se$}, Sci. Rep. \textbf{5}, 14293 (2015).

\bibitem{ele-detection-Vaklinova} K. Vaklinova, A. Hoyer, M. Burghard and K. Kern, \textit{Current-Induced Spin Polarization in Topological Insulator - Graphene Heterostructures}, Nano Lett. \textbf{16}, 2595 (2016).


\bibitem{CIS-model-Hong} S. Hong, V. Diep, S. Datta, and Y. P. Chen, \textit{Modeling Potentiometric Measurements in Topological Insulators Including Parallel Channels}, Phys. Rev. B \textbf{86}, 085131 (2012).

\bibitem{Rashba-Bianchi} M. Bianchi, D. Guan, S. Bao, J. Mi, B. B. Iversen, P. D. C. King, and P. Hofmann, \textit{Coexistence of the Topological State and a Two-Dimensional Electron Gas on the Surface of $Bi_2Se_3$}, Nat. Commun. \textbf{1}, 128 (2010).

\bibitem{Rashba-King} P. D. C. King, R. C. Hatch, M. Bianchi, R. Ovsyannikov, C. Lupulescu, G. Landolt, B. Slomski, J. H. Dil, D. Guan, J. L. Mi, E. D. L. Rienks, J. Fink, A. Lindblad, S. Svensson, S. Bao, G. Balakrishnan, B. B. Iversen, J. Osterwalder, W. Eberhardt, F. Baumberger, and Ph. Hofmann, \textit{Large Tunable Rashba Spin Splitting of a Two-Dimensional Electron Gas in $Bi_2Se_3$}, Phys. Rev. Lett. \textbf{107}, 096802 (2011).

\bibitem{Rashba-Zhu} Z.-H. Zhu, G. Levy, B. Ludbrook, C. N. Veenstra, J. A. Rosen, R. Comin, D. Wong, P. Dosanjh, A. Ubaldini, P. Syers, N. P. Butch, J. Paglione, I. S. Elfimov, and A. Damascelli, \textit{Rashba Spin-Splitting Control at the Surface of the Topological Insulator $Bi_2Se_3$}, Phys. Rev. Lett. \textbf{107}, 186405 (2011).

\bibitem{Rashba-Bahramy} M. S. Bahramy, P. D. C. King, A. de la Torre, J. Chang, M. Shi, L. Patthey, G. Balakrishnan, P. Hofmann, R. Arita, N. Nagaosa, and F. Baumberger, \textit{Emergent Quantum Confinement at Topological Insulator Surfaces}, Nat. Commun. \textbf{3}, 1159 (2012).

\bibitem{SpinVoltage-Johnson-01} M. Johnson and R. H. Silsbee, \textit{Interfacial Charge-Spin Coupling: Injection and Detection of Spin Magnetization in Metals}, Phys. Rev. Lett. \textbf{55}, 1790 (1985).

\bibitem{SpinVoltage-Johnson-02} M. Johnson and R. H. Silsbee, \textit{Coupling of Electronic Charge and Spin at a Ferromagnetic-Paramagnetic Metal Interface}, Phys. Rev. B \textbf{37}, 5312 (1988).

\bibitem{SpinDetectTheo-Schmidt} G. Schmidt, D. Ferrand, L. W. Molenkamp, A. T. Filip, and B. J. van Wees, \textit{Fundamental Obstacle for Electrical Spin Injection from a Ferromagnetic Metal into a Diffusive Semiconductor}, Phys. Rev. B \textbf{62}, R4790 (2000).

\bibitem{SpinDetectTheo-Rashba} E. I. Rashba, \textit{Theory of Electrical Spin Injection: Tunnel Contacts as a Solution of the Conductivity Mismatch Problem}, Phys. Rev. B \textbf{62}, R16267 (2000).

\bibitem{SpinDetectExp-Lou} X. Lou, C. Adelmann, S. A. Crooker, E. S. Garlid, J. Zhang, K. S. M. Reddy, S. D. Flexner, C. J. Palmstrom, and P. A. Crowell, \textit{Electrical Detection of Spin Transport in Lateral Ferromagnet-Semiconductor Devices}, Nat. Phys. \textbf{3}, 197 (2007).

\bibitem{ARPES-BSTS2-Arakane} T. Arakane, T. Sato, S. Souma, K. Kosaka, K. Nakayama, M. Komatsu, T. Takahashi, Z. Ren, K. Segawa, and Y. Ando, \textit{Tunable Dirac Cone in the Topological Insulator Bi$_{2-x}$Sb$_{x}$Te$_{3-y}$Se$_y$}, Nat. Commun. \textbf{3}, 636 (2012).

\bibitem{BSTS2-Ren} Z. Ren, A. A. Taskin, S. Sasaki, K. Segawa, and Y. Ando, \textit{Optimizing Bi$_{2-x}$Sb$_{x}$Te$_{3-y}$Se$_{y}$ Solid Solutions to Approach the Intrinsic Topological Insulator Regime}, Phys. Rev. B \textbf{84}, 165311 (2011).

\bibitem{topgate-Yang} F. Yang, A. A. Taskin, S. Sasaki, K. Segawa, Y. Ohno, K. Matsumoto, and Y. Ando, \textit{Top Gating of Epitaxial (Bi$_{1-x}$Sb$_{x}$)$_2$Te$_3$ topological insulator thin films}, Appl. Phys. Lett. \textbf{104}, 161614 (2014).

\bibitem{FringeField-Vries} E. K. de Vries, A. M. Kamerbeek, N. Koirala, M. Brahlek, M. Salehi, S. Oh, B. J. van Wees, and T. Banerjee, \textit{Towards the Understanding of the Origin of Charge-Current-Induced Spin Voltage Signals in the Topological Insulator Bi$_2$Se$_3$}, Phys. Rev. B \textbf{92}, 201102 (2015).

\bibitem{Supplemental} See Supplemental Material for details regarding the estimation of Fermi-level in BiSbTeSe$_2$ flakes, control experiment on a device without tunnel barrier, surface morphology of exfoliated BiSbTeSe$_{2}$ flakes, $R-T$ curves of the exfoliated BiSbTeSe$_{2}$ flakes and additional information about the FM tunnel contacts.

\bibitem{SignProblem-Tsymbal} E. Y. Tsymbal, O. N. Mryasov and P. R. LeClair, \textit{Spin-Dependent Tunnelling in Magnetic Tunnel Junctions}, J. Phys.: Cond. Matter \textbf{15}, R109 (2003).

\bibitem{critism-Li} P.-K. Li and I. Appelbaum, \textit{Interpreting Current-Induced Spin Polarization in Topological Insulator Surface States},  Phys. Rev. B \textbf{93}, 220404 (2016).

\bibitem{SpinHall-Seki} T. Seki, Y. Hasegawa, S. Mitani, S. Takahashi, H. Imamura, S. Maekawa, J. Nitta and K. Takanashi, \textit{Giant Spin Hall Effect in Perpendicularly Spin-Polarized FePt/Au Devices}, Nat. Mater. \textbf{7}, 125 (2008).

\bibitem{WorkFunction-Py-Saito} S. Saito and T. Maeda, \textit{Work Function of Ferromagnetic Metals and Alloys}, Vacuum (Japan) \textbf{24}, 220 (1981).

\bibitem{WorkFunction-BSTS2-Saito} D. Takane, S. Souma, T. Sato, T. Takahashi, K. Segawa, Y. Ando, \textit{Work Function of Bulk-Insulating Topological Insulator Bi$_{2-x}$Sb$_{x}$Te$_{3-y}$Se$_{y}$}, arXiv:1606.07933 (2016).


\end{thebibliography}

\begin{thebibliography}{10}

\bibitem{ARPES-BSTS2-Arakane} T. Arakane, T. Sato, S. Souma, K. Kosaka, K. Nakayama, M. Komatsu, T. Takahashi, Z. Ren, K. Segawa, and Y. Ando, \textit{Tunable Dirac Cone in the Topological Insulator Bi$_{2-x}$Sb$_{x}$Te$_{3-y}$Se$_y$}, Nat. Commun. \textbf{3}, 636 (2012).

\bibitem{ARPES-Bi2Se3-Chen} Y. L. Chen, J.-H. Chu, J. G. Analytis, Z. K. Liu, K. Igarashi, H.-H. Kuo, X. L. Qi, S.-K. Mo, R. G. Moore, and D. H. Lu, \textit{Massive Dirac Fermion on the Surface of a Magnetically Doped Topological Insulator}, Science \textbf{329}, 659 (2010).

\bibitem{LocalHall-Vries} E. K. de Vries, A. M. Kamerbeek, N. Koirala, M. Brahlek, M. Salehi, S. Oh, B. J. van Wees, and T. Banerjee, \textit{Towards the Understanding of the Origin of Charge-Current-Induced Spin Voltage Signals in the Topological Insulator Bi$_2$Se$_3$}, Phys. Rev. B \textbf{92}, 201102 (2015).

\end{thebibliography}
\end{document}